\documentclass[a4paper11pt]{article}

\usepackage{jheppub}
\usepackage[T1]{fontenc}
\usepackage{graphicx, hyperref}
\usepackage{psfrag}
\usepackage{textcomp}

\newcommand{\prd}{Phys.~Rev.~D}
\newcommand{\pre}{Phys.~Rev.~E}

\title{Thermodynamics, stability and Hawking--Page transition of Kerr black holes from R\'enyi statistics}

\author[a,b,1]{Viktor G.~Czinner\note{Corresponding author.}}
\author[c]{and Hideo Iguchi}

\affiliation[a]{Multidisciplinary Center for Astrophysics \& Department of Physics,
Instituto Superior T\'ecnico, University of Lisbon, Avenida Rovisco Pais 1, 1049-001 Lisboa, Portugal}
\affiliation[b]{HAS Wigner Research Centre for Physics, H-1525 Budapest, P.O.~Box 49, Hungary}
\affiliation[c]{Laboratory of Physics, College of Science and Technology, Nihon University,
274-8501 Narashinodai, Funabashi, Chiba, Japan}

\emailAdd{viktor.czinner@tecnico.ulisboa.pt}
\emailAdd{iguchi.h@phys.ge.cst.nihon-u.ac.jp}

\abstract{
Thermodynamics of rotating black holes described by the R\'enyi formula as equilibrium 
and zeroth law compatible entropy function is investigated. We show that similarly to 
the standard 
Boltzmann approach, isolated Kerr black holes are stable with respect to axisymmetric 
perturbations in the R\'enyi model. On the other hand, when the black holes are 
surrounded by a bath of thermal radiation, slowly rotating black holes can also be 
in stable equilibrium with the heat bath at a fixed temperature, in contrast to
the Boltzmann description. For the question of possible phase transitions 
in the system, we show that a Hawking--Page transition and a first order 
small black hole/large black hole transition occur, analogous to the 
picture of rotating black holes in AdS space. These results confirm the similarity 
between the R\'enyi-asymptotically flat and Boltzmann-AdS approaches to black hole
thermodynamics in the rotating case as well. We derive the relations between the 
thermodynamic parameters based on this correspondence.}

\begin{document}
\maketitle
\flushbottom

\section{Introduction}

Gravitational phase transitions, in particular the ones connected to black hole thermodynamics, 
are essential constituents of many open problems in modern theoretical physics. The Hawking--Page 
phase transition \cite{Hawking:1982dh} of black holes in anti-de Sitter space is 
one of the most important ones due to its role in the AdS/CFT correspondence 
\cite{Maldacena:1997re,Witten:1998qj} and also in related phenomena of confinement/deconfinement 
transitions at finite temperature in various gauge theories \cite{MMT1,MMT2}. Because of  
the different background geometry, asymptotically flat black holes have different stability 
properties than AdS ones, and in the standard black hole thermodynamic picture 
\cite{Bekenstein:1973ur,Bardeen:1973gs,Hawking:1974sw,Hawking:1976de}, they mostly tend to be 
unstable for any large masses when surrounded by an infinite bath of thermal radiation. 
A Hawking--Page transition does not occur under these conditions, and a cosmic 
black hole nucleation is not present in asymptotically flat spacetimes. 
Apart from the gravity interest, the above phenomenon is interesting from a thermodynamic 
viewpoint as well, and for a clear understanding of the physics behind, 
the underlying theory of black hole thermodynamics is also necessary to be well understood. 
In the past 40 years, after the 
foundations of the standard thermodynamic theory of black holes
\cite{Bekenstein:1973ur,Bardeen:1973gs,Hawking:1974sw,Hawking:1976de},
numerous achievements have been made in the field. In spite of the
active research and successes however, there still are some unsettled and important issues which 
could not be resolved satisfactorily so far. From a classical thermodynamic perspective,
one of the most interesting ones is the nonextensive nature of black holes and the corresponding
problem of thermodynamic stability. 

A basic group of physical quantities in classical thermodynamics is the group called 
extensive variables $X$ (like energy, entropy, etc.), where it is assumed 
that these quantities are \emph{additive} for composition, i.e.~$X_{12}=X_1+X_2$ when 
thermodynamic systems are joined together \cite{PhysRevE.83.061147}. On the other hand, 
it is also customary to assume that these quantities characterize the system down to the 
smallest scales \cite{Mackey}, i.e.~when working with finite densities:
\begin{equation}
\varrho_X = \lim_{n\rightarrow\infty}\frac{1}{n}\sum_{i=1}^n X_i < \infty,
\end{equation}
where the system is divided into $n$ different parts. This property is called 
\emph{extensivity}. The two properties: 
additivity and extensivity are not equivalent. An additive quantity is 
extensive, but extensive quantities can be nonadditive too \cite{T2,B1}.
Black holes are very peculiar creatures in this respect because
they cannot be described as the union of some constituent subsystems 
which are endowed with their own thermodynamics, and therefore black holes
are nonextensive objects. 

Looking from a different perspective, phenomenological thermodynamics of 
macroscopic objects has a well understood theory from statistical physics 
where the macroscopic properties of a given body (described by the thermodynamic 
parameters e.g.~total energy, entropy, temperature, etc.)~can be uniquely 
obtained from the microscopic description of the system. Standard statistical 
descriptions, on the other hand, usually assume that long-range type 
interactions are negligible, i.e.~that the (linear) size 
of the system in question is much lager than the range of the relevant 
interaction between the elements of the system. Under these conditions 
the standard local notions of mass, energy and other extensive quantities 
are well defined, and by applying the additive (and therefore extensive) 
Boltzamann--Gibbs formula: $S_{BG}=-\sum p_i\ln p_i$, for defining the 
system's entropy function, the classical thermodynamic description is 
recovered in the macroscopic limit. 

In the presence of strong gravitational fields however, and in particular 
when black holes are considered, the assumption of negligible long-range 
type interactions can not be hold, and consequently the usual definition 
of mass and other extensive quantities is not possible locally. Nonlocality 
is indeed a fundamental feature of general relativity, and corresponding 
nonextesive thermodynamic phenomena have been known in cosmology and 
gravitation theory for a long time
(see e.g.~\cite{Landsberg1984, cg1,cg2,cg3,cg4,cg5,cg6,cg7,cg8,cg9} 
and references therein). In fact, even as early as 1902, Gibbs already pointed 
out in his statistical mechanics book \cite{Gibbs}, that systems with divergent 
partition function lie outside the validity of Boltzamann--Gibbs theory. He 
explicitly mentions gravitation as an example (see e.g.~\cite{Tsallis:2012js} 
for more details). Therefore, the standard Boltzamann--Gibbs statistics may not be the 
best possible choice for defining the entropy function in strongly gravitating 
systems, and other statistical approaches, which could also take into account 
the long-range type property of the relevant interaction (i.e.~gravitation) and 
the nonextensive nature of the problem, are also relevant and important to study.

The nonextensive nature of the Bekenstein-Hawking entropy of black hole 
event horizons has been noticed \cite{Davies:1978mf} very early on after the 
thermodynamic theory of black holes had been formulated 
\cite{Bekenstein:1973ur,Bardeen:1973gs,Hawking:1974sw,Hawking:1976de}, and 
the corresponding thermodynamic and stability problem has been investigated 
several times with various approaches (see e.g.~\cite{landsberg1980entropies,bishop1987thermodynamics,Landsberg1984,
pavon1986some,1993Natur.365..103M,Gour:2003pd,Oppenheim:2002kx,Pesci:2006sb,Aranha:2008ni,Tsallis:2012js} 
and references therein). 
The general theory of nonadditive thermodynamics has also advanced significantly in the past 
few decades (see e.g.~\cite{T2,Landsberg} and references therein), and it has been shown,
that by relaxing the additivity requirement in the axiomatic approach to the entropy
definition (given by Shannon \cite{Shannon} and Khinchin \cite{Khinchin}) to the 
weaker \emph{composability} requirement, new possible functional forms of the entropy 
may arise \cite{Tempesta}. As a consequence, there exist certain parametric extensions 
of the Boltzamann--Gibbs statistical entropy formula, which seem to be more appropriate to 
describe systems with long-range type interactions. One such statistical entropy 
definition has been proposed by Tsallis \cite{Tsallis:1987eu} as:
\begin{equation}\label{TsS}
 S_T=\frac{1}{1-q}\sum_i(p^q_i-p_i),
\end{equation}
where $p_i$ are the probabilities of the microscopic states of the system, and 
$q\in\mathbb{R}$ is the so-called nonextensivity parameter. In the limit  
of $q\rightarrow 1$, $S_T$ reproduces the standard Boltzamann--Gibbs result, however,
in the case when $q\neq 0$, the Tsallis entropy is not additive, and the parameter 
can be attributed to measure the effects of non-localities in the system. 
The $q$-parameter is usually constant in different physical situations, and its 
explicit value is part of the problem to be solved. 

The Tsallis statistics to the black hole problem has been investigated with various
approaches (see e.g.~\cite{Tsallis:2012js} and references therein), however it has 
been a long-standing problem in nonextensive thermodynamics that nonadditive entropy 
composition rules (in general) 
can not be compatible with the most natural requirement of thermal equilibrium in the 
system \cite{PhysRevE.83.061147}. They usually don't satisfy the zeroth law of 
thermodynamics, which requires the existence of a well defined, unique, empirical 
temperature in thermal equilibrium which is constant all over the system. 
For resolving these issues, Bir\'o and V\'an developed a method \cite{PhysRevE.83.061147}, 
called the "\emph{formal logarithm approach}", which maps the 
original, nonadditive entropy composition rule of a given system to an additive 
one by a simple transformation. This procedure results a new, but also well defined 
entropy function for the system, which in turn, also satisfies both the equilibrium 
and the zeroth law compatibility requirements of thermodynamics. In case of the
nonextensive Tsallis statistics, this new entropy turns out to be the well known 
R\'enyi formula \cite{renyi1959dimension,renyi1970probability}, defined as
\begin{equation}\label{Srenyi}
S_R=\frac{1}{1-q}\ln\sum_ip^q_i,
\end{equation}
which had been proposed earlier by the Hungarian mathematician Alfr\'ed R\'enyi in 1959 
\cite{renyi1959dimension}.

Recently, motivated by the nonextensive and nonlocal nature of black hole thermodynamics,
we proposed and studied an alternative approach to the black hole entropy problem
\cite{Biro:2013cra}. In this model, in order to satisfy both the equilibrium the zeroth law 
compatibility, instead of the Tsallis description, we considered its formal logarithm, the 
R\'enyi statistics (\ref{TsS}) to describe the thermodynamic entropy of black hole event 
horizons. The explicit details of this approach is presented in the next section, and by 
applying the R\'enyi model to Schwarzschild black holes \cite{Biro:2013cra}, we found that 
the temperature-horizon radius relation of the black hole has the 
same form as the one obtained from a black hole in anti-de Sitter space by using 
the original Boltzamann-Gibbs statistics. In both cases the temperature has a minimum. 
By using a semi-classical estimate on the horizon radius at this minimum, we 
obtained a Bekenstein bound \cite{Bekenstein:1980jp} for the $q$-parameter 
value in the R\'enyi entropy of micro black holes ($q \geq 1 + 2/\pi^2$), which 
was surprisingly close to other $q$-parameter fits from very distant and unrelated 
physical phenomena, e.g.~cosmic ray spectra \cite{Bek1,Bek2}, and power-law 
distribution of quarks coalescing to hadrons in high energy accelerator 
experiments \cite{BU}.

Besides the statistical approach, another fundamental problem of applying standard 
thermodynamic methods to black holes arising from the question of stability. In 
ordinary thermodynamics of extensive systems, the local thermodynamic stability 
(defined as the Hessian of the entropy has no positive eigenvalues) is linked to 
the dynamical stability
of the system. This stability criteria, however, strongly relies on the additivity 
of the entropy function, which is a property that clearly does not hold for black 
holes. The simplest example of this discrepancy is the Schwarzschild black hole 
which is known to be perturbatively stable but has a negative specific heat
(positive Hessian). Black hole phase transitions are also strongly related to the 
stability properties of the system (in particular the Hawking--Page transition), 
and since the standard methods are not reliable in nonextensive thermodynamics, 
one has to be very careful when considering stability and phase transitions in 
strongly gravitating systems.

Avoiding the complications arising from the Hessian approach to the stability problem of 
black holes, an alternative technique was proposed in a series of paper by Kaburaki 
et al.~\cite{Kaburaki:1993ah,Katz:1993up,kaburaki1996critical}. In these works 
the so-called "Poincar\'e turning point method of stability" \cite{poincare1885equilibre}
has been applied to the problem, which is a topological approach and 
does not depend on the additivity of the entropy function. More recently this method has 
also been used to study critical phenomena of higher dimensional black holes and black 
rings \cite{Arcioni:2004ww} and to determine the conditions of stability for equilibrium 
configurations of charged black holes surrounded by quintessence \cite{AzregAinou:2012hy}.
In section \ref{stab} we present an overview of this method.


By considering the R\'enyi model in the black hole problem, we also investigated 
the thermodynamic stability question of Schwarzschild black holes \cite{Czinner:2015eyk}. 
First we considered the question of pure, isolated black holes in the microcanonical approach, 
and showed that these configurations are stable against spherically symmetric perturbations, 
just like in the Boltzmann picture. However, in considering the case when the black holes are 
surrounded by a heat bath in the canonical treatment, we found that -- in contrast to the 
Boltzmann approach -- Schwarzschild black holes can be in stable equilibrium with thermal 
radiation at a fixed temperature. This results a stability change at a certain value of 
the mass-energy parameter of the black hole which belongs to the minimum temperature solution. 
Black holes 
with smaller masses are unstable in this model, however larger black holes become stable. 
These findings are essentially identical to the ones obtained by Hawking and Page in AdS 
space within the standard Boltzmann entropy description \cite{Hawking:1982dh}. According 
to this similarity, we also analyzed the question of a possible phase transition in the 
canonical picture and found that a Hawking--Page black hole phase transition occurs in a 
very similar fashion as in AdS space in the Boltzamann statistics. We showed that the 
corresponding critical temperature depends only on the $q$-parameter of the R\'enyi formula, 
just like it depends only on the curvature parameter in AdS space. For the stability analysis 
we considered both the Poincar\'e and the Hessian methods. The latter one could also be 
applied since the R\'enyi entropy is additive for composition (see the next section), and 
therefore the standard stability analysis is also reliable in this case. Both approaches 
confirmed the same stability results.

These findings might have some relevant consequences in black hole physics. In particular, 
if an effective physical model could be constructed on how to compute the 
$q$-parameter value for the R\'enyi entropy of black holes (or other strongly gravitating 
systems) in order to parametrize the non-local effects of the gravitational field, 
the R\'enyi statistics can provide a well behaving and additive entropy description of the 
system which is also compatible with the requirements of equilibrium and the zeroth 
law of thermodynamics. Similar considerations have been applied recently to describe 
the relative information entropy measure inside compact domains of an inhomogeneous 
universe \cite{CzM}, where an explicit geometric model has been proposed to compute the 
$q$-parameter of the R\'enyi entropy in order to measure the effects of the gravitational
entanglement problem. 

In the case of black hole thermodynamics, we showed 
that large, asymptotically flat, Schwarzschild black holes can be in stable equilibrium 
with a thermal heath bath in the R\'enyi picture, and a Hawking--Page phase transition
can occur in the system. This result offers a possible explanation for the problem 
of cosmic black hole nucleation in the early universe, and this mechanism might be the 
origin of large or super massive black holes that can be found in most galaxy centers. 
Many other interesting consequences can be deduced from the R\'enyi approach, but in this 
work we aim to achieve a more modest goal.
In the present paper, by extending our previous investigations in the problem, we study the 
thermodynamic, stability and phase transition properties of Kerr black holes within the 
R\'enyi model and analyze whether similar results can be obtained to what 
we have found in the Schwarzschild case. In this analysis the turning point method is 
applied to the stability problem in both the microcanonical and canonical ensembles, and 
we will show that stability changes appear in the latter case, which suggests that a 
Hawking--Page transition and a first order small black hole/large black hole phase 
transition occur in the system, similar to the one observed for charged 
and rotating black holes in AdS space \cite{Chamblin:1999tk,Chamblin:1999hg,
Caldarelli:1999xj,Tsai:2011gv,Altamirano:2014tva}. This result provides a correspondence 
between the Kerr--R\'enyi and the Kerr-AdS--Boltzmann pictures, analogous to the one 
we reported in the Schwarzschild problem \cite{Biro:2013cra,Czinner:2015eyk}.

The plan of the paper is as follows. In Sec.~\ref{sec:Renyi} we discuss the foundations and 
motivation of the R\'enyi approach arising from nonextensive thermodynamics to the 
black hole problem. In Sec.~\ref{sec:Kerr} we introduce the Kerr solution and calculate its 
thermodynamic quantities within the R\'enyi statistics. In Sec.~\ref{sec:Stability} we 
investigate the thermodynamic stability problem of Kerr black holes in the R\'enyi model
by the Poincar\'e turning point method both in the microcanonical and canonical treatments. 
We also discuss the question of possible phase transitions in this section. In Sec.~\ref{sec:Kerr-AdS} 
the thermodynamic stability problem of Kerr-AdS black holes in the standard Boltzmann case is also 
presented by the turning point method, and the correspondence between the Kerr--R\'enyi and the 
Kerr-AdS--Boltzmann approaches is discussed. In Sec.~\ref{sec:summary} we summarize our results and 
draw our conclusions. Throughout this paper we use units such as $c=G=\hbar=k_B=1$.

\section{R\'enyi approach from nonadditive thermodynamics} 
\label{sec:Renyi}

By replacing the additivity axiom to the weaker composability in the Shannon--Khinchin axiomatic
definition of the entropy function, new type of entropy expressions arise. The composability axiom
asserts, roughly speaking, that the entropy $S_{12}$ of a compound system consisting of two independent
systems should be computable only in terms of the individual entropies $S_{1}$ and $S_{2}$. This
means that there is a function $f(x,y)$ such that 
\begin{equation}
 S_{12} = f (S_1, S_2),
\end{equation}
for any independent systems. This property is of fundamental importance, since it implies that an
entropic function is properly defined on macroscopic states of a given system, and it can be computed
without having any information on the underlying microscopic dynamics. Composability is a key feature 
to ensure that the entropy function is physically meaningful.
In a recent paper \cite{abe2001general}, based on the concept of composability alone, Abe derived the 
most general functional form of those nonadditive entropy composition rules that are compatible 
with homogeneous equilibrium. Assuming that $f(S_1, S_2)$ is a $C^2$ class symmetric function, 
Abe showed that the most general, equilibrium compatible composition rule takes the form
\begin{equation}\label{eq:Abe}
H_{\lambda}(S_{12})=H_{\lambda}(S_1)+H_{\lambda}(S_2)+\lambda H_{\lambda}(S_1)H_{\lambda}(S_2), 
\end{equation}
where $H_{\lambda}$ is a differentiable function of $S$ and $\lambda\in\mathbb{R}$ is a constant 
parameter. Later on, this result has been extended to non-homogeneous systems as well 
\cite{PhysRevE.83.061147}, where not only the entropy, but the energy function is also considered 
to be nonadditive.

The simplest and perhaps the most well-known nonadditive entropy composition rule can 
be obtained from (\ref{eq:Abe}) by setting $H_{\lambda}(S)$ to be the identity function, 
i.e.~$H_{\lambda}(S)=S$. In this case Abe's equation becomes  
\begin{equation}\label{eq:tsallis_composition}
S_{12}=S_1+S_2+\lambda S_1 S_2,
\end{equation}
which results the familiar Tsallis composition rule with $\lambda=1-q$ \cite{Tsallis:1987eu}, 
and the corresponding entropy definition is given in (\ref{TsS}). The nonextensive
Tsallis statistics is widely investigated in many research fields from natural to
social sciences, an updated bibliography on the topic can be found in \cite{Tsallis-bib}.
This approach has also been studied in the problem of black hole thermodynamics 
(see e.g.~\cite{Tsallis:2012js} and references therein), and our 
starting point in considering a more general entropy definition for black holes than 
the one based on the Boltzamann-Gibbs statistics is also the Tsallis formula.

Generalized, nonadditive entropy definitions has been investigated in various problems
from high energy physics \cite{B1} to DNA analysis \cite{DNA}, and it has been a longstanding 
problem that the zeroth law of thermodynamics (i.e.~the existence of a well defined temperature 
function in thermal equilibrium) cannot be compatible with nonadditive entropy composition rules. 
A possible resolution to this problem has been proposed recently by Bir\'o and V\'an in
\cite{PhysRevE.83.061147}, where they developed a formulation to determine the most 
general functional form of those nonadditive entropy composition rules that are compatible 
with the zeroth law of thermodynamics. They found that the general form is additive for 
the \emph{formal logarithms} of the original quantities, which in turn, also satisfy the familiar 
relations of standard thermodynamics. In particular, for homogeneous systems, they showed 
that the most general, zeroth law compatible entropy function takes the form 
\begin{equation}\label{eq:formallog}
L(S)=\frac{1}{\lambda}\ln[1+\lambda H_{\lambda}(S)],
\end{equation}
which is additive for composition, i.e.,
\begin{equation}
L(S_{12})=L(S_{1})+L(S_{2}),
\end{equation}
and the corresponding zeroth law compatible temperature function can be obtained as
\begin{equation}
\frac{1}{T}=\frac{\partial L(S(E))}{\partial E}, 
\end{equation}
where $E$ is the energy of the system.

In the case of the Tsallis statistics, it is easy to show that by taking the formal logarithm 
(\ref{eq:formallog}) of the Tsallis entropy (\ref{TsS}), i.e.
\begin{equation}
L(S_T)=\frac{1}{1-q}\ln\left[1+(1-q)S_T\right] \equiv S_R,
\end{equation}
the R\'enyi expression (\ref{Srenyi}) is reproduced, which, unlike the Tsallis formula, is 
additive for composition. In the limit of $q\rightarrow 1$ ($\lambda \rightarrow 0$), both 
the Tsallis- and the R\'enyi entropies recovers the standard Boltzmann-Gibbs description.

According to these results, in the present paper, in order to describe the non-Boltzamannian 
nature of Kerr black holes, we consider the Tsallis statistics as the simplest, nonadditive, 
parametric but equilibrium compatible extension of the Boltzamann-Gibbs theory which also
satisfies Abe's formula. On the other hand, in order to satisfy the zeroth law of thermodynamics, 
we follow the formal logarithm method of Bir\'o and V\'an, and rather the Tsallis 
description, we consider the R\'enyi entropy for the thermodynamics of the problem. Since the 
R\'enyi definition is additive, it satisfies all laws of thermodynamics, and compared to the 
Boltzmann picture, it has the advantage of having a free parameter which can be accounted to 
describe the effects of nonlocality in our approach. The thermodynamics of Schwarzschild black 
holes in this model has been studied in \cite{Biro:2013cra}, and the corresponding stability 
problem has been investigated in \cite{Czinner:2015eyk}.

\section{Kerr black holes}
\label{sec:Kerr}
\begin{figure}
  \noindent\hfil\includegraphics[scale=0.7,angle=0]{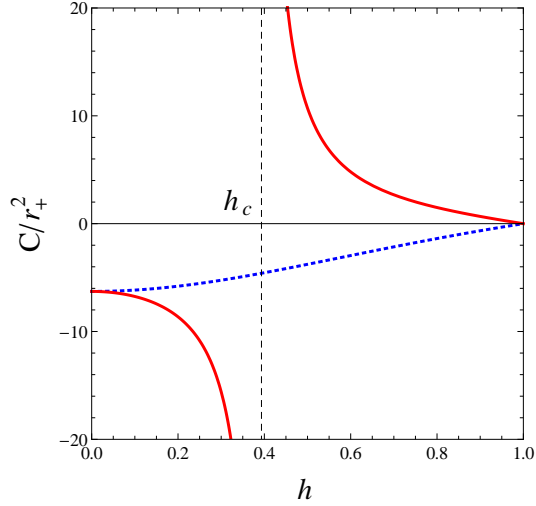}
  \caption{Plots of the heat capacities $C_J$ (red solid line) and $C_\Omega$
  (blue dotted line) against $h$. $C_J$ diverges at $h_c$ where a pole occurs.}
 \label{fig:heat_capacity_Kerr}
\end{figure}

The spacetime metric that describes the geometry of a rotating black hole is given
by the Kerr solution
\begin{eqnarray}
 ds^2 & = & -dt^2 + \frac{2 M r}{\Sigma} \left( dt -a \sin^2 \theta d \phi \right)^2 
 + \frac{\Sigma}{\Delta} dr^2 + \Sigma d \theta^2  \nonumber \\
 & &+ (r^2 +a^2) \sin \theta d \phi^2\ ,
\end{eqnarray}
where
\begin{equation}
 \Sigma = r^2 + a^2 \cos^2 \theta , ~~~ \Delta = r^2 + a^2 - 2Mr .
\end{equation}
Here, $M$ is the mass-energy parameter of the black hole and $a$ is its rotation parameter.
The thermodynamic quantities of a Kerr black hole can be expressed in terms of its 
horizon radius $r_{+} = M + \sqrt{M^2 - a^2}$, which is defined by taking $\Delta = 0$.
The Hawking temperature of the black hole horizon is
\begin{equation}
 T_H = \frac{1}{2\pi}\left[ \frac{r_{+}}{r_{+}^2 + a^2} - \frac{1}{2 r_{+}} \right],
\end{equation}
the Bekenstein-Hawking entropy is 
\begin{equation}
 S_{BH} = \pi (r_{+}^2 + a^2),
\end{equation}
the angular momentum of the black hole is
\begin{equation}
 J = \frac{a}{2 r_{+}}(r_{+}^2 + a^2),
\end{equation}
the angular velocity of the horizon is
\begin{equation}
 \Omega = \frac{a}{r_{+}^2 + a^2},
\end{equation}
and the mass-energy parameter can also be written as 
\begin{equation}
 M = \frac{r_{+}^2 + a^2}{2 r_{+}}.
\end{equation}
The heat capacity at constant angular velocity is given by
\begin{equation}
 C_{\Omega} = T_H \left( \frac{\partial S_{BH}}{\partial T_H} \right)_\Omega 
 = \frac{2 \pi r_{+}^2(a^2 - r_{+}^2)}{r_{+}^2 +a^2},
\end{equation}
and the heat capacity at constant angular momentum is
\begin{equation}
 C_J = T_H \left( \frac{\partial S_{BH}}{\partial T_H} \right)_J 
 = \frac{2 \pi (r_{+}^2 -a^2)(r_{+}^2 + a^2)^2}{3 a^4 + 6 r_{+}^2 a^2 - r_{+}^4}\ .
\end{equation}
$C_\Omega$ and $C_J$ can be written in simpler forms if
we normalize them by $r_{+}^2$, i.e.
\begin{equation}
 \frac{C_{\Omega}}{r_{+}^2} = - \frac{2 \pi (1 - h^2)}{h^2 + 1 }
\end{equation}
and
\begin{equation}
 \frac{C_{J}}{r_{+}^2} =  \frac{2 \pi (1 - h^2 )(h^2 + 1)^2}{3 h^4 + 6 h^2 - 1},
\end{equation}
were we also introduced the normalized rotation parameter $h$ {\cite{okamoto1990thermodynamical}} 
as \[h \equiv \frac{|a|}{r_{+}} .\]
The $r_{+}$ horizon radius exists only for $|a| \le M$, which corresponds to $0 \le h \le 1$.
The $h=0$ value describes a Schwarzschild black hole, while the limiting value $h=1$ belongs 
to the extreme Kerr black hole case. The heat capacities as functions of $h$ are plotted on 
Fig.~\ref{fig:heat_capacity_Kerr}. It can be seen that $C_\Omega$ is negative for $ 0\le h<1$ 
and $C_J$ diverges at $h_c = \sqrt{\frac{2}{3}\sqrt{3} - 1}$, where a pole occurs. 
$C_J$ is negative for $h<h_c$ and positive for $h>h_c$ values. The heat capacities 
coincide at the limit values $h=0$ and 1.

The R\'enyi entropy function of black holes can be obtained by taking the formal 
logarithm of the Bekenstein-Hawking entropy, which in our non-Boltzmannian approach 
follows the nonadditive Tsallis statistics. The physical meaning of the $\lambda$ 
parameter is connected to the nonextensive and nonlocal nature of the problem, and 
the R\'enyi entropy of a general black hole in this picture is given by
\begin{equation}
 S_R = \frac{1}{\lambda} \ln (1 + \lambda S_{BH}).
\end{equation}
The zeroth law compatible R\'enyi temperature is then defined as 
\begin{equation}
  T_R = \frac{1}{\partial S_R / \partial M} = T_H ( 1 + \lambda S_{BH} ).
\end{equation}

\begin{figure}
  \noindent\hfil\includegraphics[scale=0.7,angle=0]{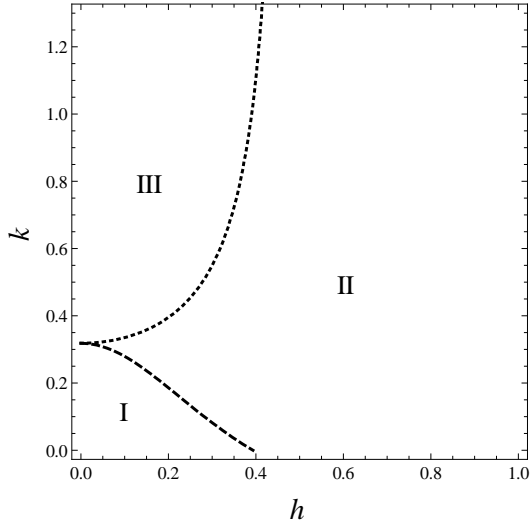}
  \caption{Phase diagram of Kerr black holes with R\'enyi entropy. 
  On the dashed curve $C_{JR}$ diverges while on the dotted curve 
  $C_{\Omega R}$ diverges. In region I, $C_{\Omega R}<0$ and 
  $C_{JR} <0$, in region II, $C_{\Omega R}<0$ and $C_{JR} >0$, while 
  in region III, $C_{\Omega R}>0$ and $C_{JR} >0$.}
 \label{fig:heat_capacity}
\end{figure}

For the case of a Kerr black hole, the R\'enyi entropy and the corresponding temperature 
take the forms
\begin{equation}
 \label{eq:Renyi_entropy}
 S_R = \frac{1}{\lambda} \ln ( 1 + \pi \lambda (r_{+}^2 + a^2) )
\end{equation}
and
\begin{equation}
 \label{eq:Renyi_temperature}
 T_R = \frac{(1 + \pi \lambda (r_{+}^2 + a^2))(r_{+}^2 - a^2)}{4 \pi r_{+} (r_{+}^2 + a^2)}.
\end{equation}
The heat capacities can be obtained as
\begin{equation}
 C_{R} = T_R \left( \frac{\partial S_R}{\partial T_R} \right) 
 = \frac{C_{BH}}{1 + \lambda(S_{BH} + C_{BH})},
\end{equation}
where $C_{BH} = T_H \left( \frac{\partial S_{BH}}{\partial T_H} \right)$.
The heat capacity at constant angular velocity is then 
\begin{equation}
 C_{\Omega R} = \frac{2 \pi r_{+}^2(a^2 - r_{+}^2)}{r_{+}^2 + a^2 
 + \pi \lambda (a^4 - r_{+}^4 + 4 a^2 r_{+}^2)},
\end{equation}
while the heat capacity at constant angular momentum takes the form
\begin{equation}
 C_{JR} = \frac{2 \pi (r_{+}^2 - a^2)(a^2 + r_{+}^2)^2}{3 a^4 +6 a^2 r_{+}^2 - r_{+}^4 
 + \pi \lambda (a^2 + r_{+}^2)(a^4 +6 r_{+}^2 a^2 +r_{+}^4)}.
\end{equation}
$C_{\Omega R}$ and $C_{JR}$ can also be written in the simpler, normalized 
forms as before, i.e.
\begin{equation}
 \frac{C_{\Omega R}}{r_{+}^2} = \frac{2 \pi (h^2 -1)}{h^2 + 1 + \pi k (h^4+ 4 h^2 -1)},
\end{equation}
and
\begin{equation}
 \frac{C_{JR}}{r_{+}^2} = - \frac{2 \pi (h^2 - 1)(h^2 + 1)^2}{3 h^4 + 6 h^2 - 1 
 + \pi k (h^2 + 1)(h^4 + 6 h^2 + 1)},
\end{equation}
where we also introduced the parameter 
\[k = \lambda r_{+}^2.\]
The heat capacities change their sign depending on the parameter values as plotted on 
Fig.~\ref{fig:heat_capacity}.

\begin{figure}
  \noindent\hfil\includegraphics[scale=0.7,angle=0]{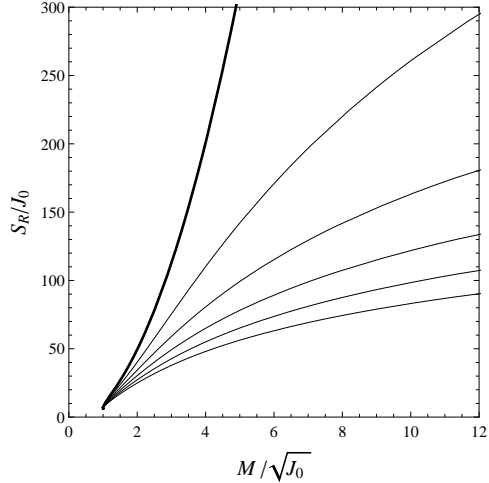}
  \caption{Plots of the R\'enyi entropy as a function of the mass-energy parameter at 
  fixed $J = J_0$ for the parameter values $\lambda J_0 = 0.05$, 0.04, 0.03, 0.02 and 
  0.01 starting from the bottom curve respectively. The top, bold curve belongs to the 
  standard Boltzmann entropy of the Kerr black hole.}
 \label{fig:MS_Kerr}
\end{figure}

For fixed angular momentum, both the R\'enyi and the 
Boltzmann entropies of a Kerr black hole are monotonically increasing functions 
of the mass parameter (Fig.~\ref{fig:MS_Kerr}). An important difference however, 
is that while the {standard Boltzmann entropy is asymptotically convex} 
(being proportional to $M^2$ as approaching the static Schwarzschild 
solution in the large $M$ limit), {{the R\'enyi entropy is asymptotically concave, since it increases only logarithmically.}}

\begin{figure}
  \noindent\hfil\includegraphics[scale=0.7,angle=0]{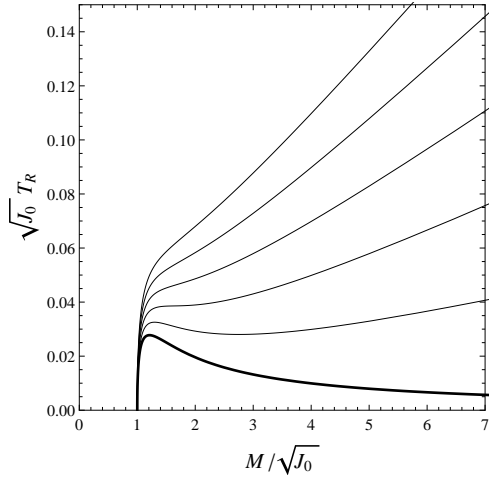}
  \caption{Plots of the R\'enyi temperature against the mass-energy parameter 
  of Kerr black holes at fixed $J = J_0$ with $\lambda J_0 = 0.05$, 0.04, 0.03, 
  0.02 and 0.01 parameter values starting from the top respectively. The bottom, 
  bold curve belongs to the standard Hawking temperature of the black hole in the 
  Boltzmann model.}
 \label{fig:MT_Kerr}
\end{figure}

On Fig.~\ref{fig:MT_Kerr}, we also plotted the temperature-energy relations for fixed 
angular momentum $J = J_0$. As it is well known, there is a maximum temperature in the 
case of the standard Boltzmann approach. In the smaller  mass (low entropy) region of 
the $T_H(M)$ curve, the heat capacity $C_J$ is positive, while at larger masses 
(high entropy) region it is negative. The two regions correspond to the phases of 
$h>h_c$ and $h<h_c$ on Fig.~\ref{fig:heat_capacity_Kerr}. In the R\'enyi approach, 
the $T_R(M)$ curve has different behavior depending on the actual parameter values
of the angular momentum and $\lambda$. It has 
a local maximum and a local minimum when $\lambda J_0$ is smaller than some critical 
value. In this case three black holes with different masses can coexist at a given 
temperature. The $C_{JR}$ heat capacity is negative in the phase between the two 
local extrema and positive otherwise (see also Fig.~\ref{fig:hc_J0} in Sec.~\ref{sec:Stability}).
Above the critical value of $\lambda J_0$, the local extrema disappear and the R\'enyi 
temperature becomes a monotonically increasing function of the energy. From these 
properties one can expect a stability change and a thermodynamic phase transition of Kerr 
black holes in the canonical ensemble. We will investigate this question in the next section.

\section{Stability analysis}\label{stab}

Kaburaki {\it et al.}~\cite{Kaburaki:1993ah} 
have shown that isolated Kerr black holes are stable against axisymmetric perturbations in 
the standard picture of black hole thermodynamics. They also investigated the effects of thermal 
radiation around Kerr black holes from the standpoint of a canonical ensemble. They concluded that 
slowly rotating black holes (just like Schwarzschild black holes) are unstable in a heat bath,
but rapidly rotating holes are less unstable and may even be stable.

\label{sec:Stability}
As we pointed out, the standard thermodynamic stability criteria based 
on the Hessian analysis fails when it is applied to black holes because 
of the nonadditivity of the entropy function, and also of the impossibility to 
define extensive thermodynamic quantities (see e.g.~\cite{Arcioni:2004ww}).
These problems can be mainly avoided by the Poincar\'e turning point method 
\cite{poincare1885equilibre}, which can divide a linear series of equilibrium 
into subspaces of unstable and less unstable states of equilibria solely from 
the properties of the equilibrium sequence without solving any eigenvalue equation.
The method has been widely applied to problems in astrophysical and gravitating systems, 
in particular for the study of the thermodynamic stability of black holes in four 
and also in higher dimensions \cite{Kaburaki:1993ah,Katz:1993up,kaburaki1996critical,
Arcioni:2004ww,AzregAinou:2012hy}. A well-organized review on the method has been given 
by Arcioni and Lozano-Tellechea in \cite{Arcioni:2004ww}, which we will mainly 
follow in the next brief description. Formal proofs of the results can be found in the 
previous works of Katz \cite{1978MNRAS.183..765K,1979MNRAS.189..817K} and Sorkin 
\cite{Sorkin:1981jc}, and a more intuitive interpretation has also been given in 
\cite{1994PhLA..185...21K}.

\begin{figure}
  \noindent\hfil\includegraphics[scale=0.25,angle=0]{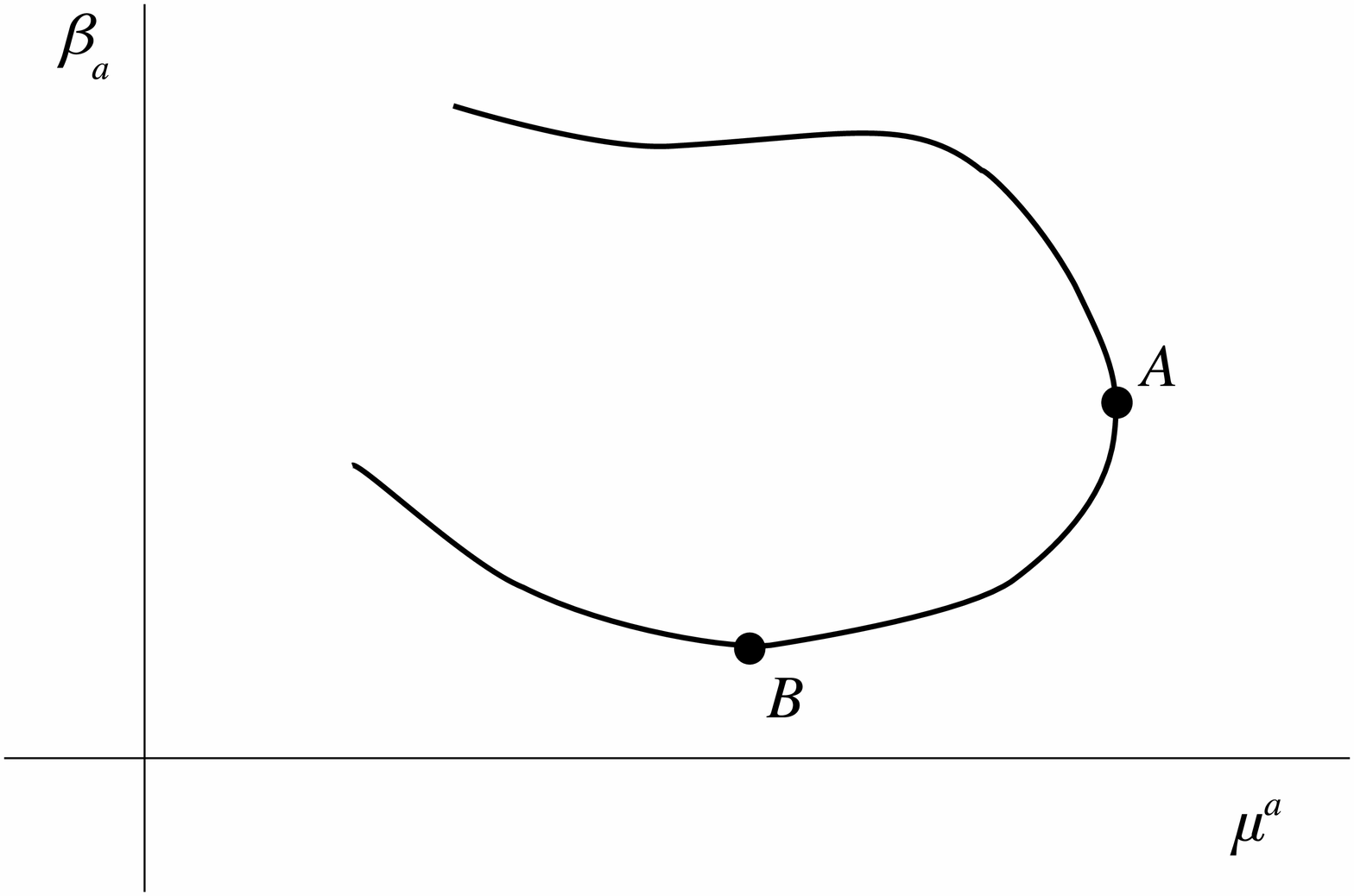}
  \caption{Example of stability curves. A conjugate variable $\beta_a$ against a control 
  parameter $\mu_a$ is plotted. The point $A$ is a turning point, where a stability change 
  can occur. At point $B$ a stability change doesn't occur even if the slope 
  changes its sign there.}
 \label{fig:turning}
\end{figure}

Suppose $\mathcal{M}$ is the space of possible configurations of a system in 
which a given configuration is specified by a point $X$. The set of independent thermodynamic 
variables ${\mu^i}$ specifies a given ensemble. Let $Z$ be the corresponding Massieu function.
Both ${\mu^i}$ and $Z$ are functions on  $\mathcal{M}$. Equilibrium states occur at points in 
$\mathcal{M}$ which are extrema of $Z$ under displacements $dX$ for which $d \mu^i = 0$. 
At equilibrium the Massieu function depends only on the values of the thermodynamic variables 
$\mu^i$. We can define conjugate variables $\beta_i$ such that
\begin{equation}
 dS = \beta_i d \mu^i
 \label{eq:dS}
\end{equation}
for all displacements $dX$.
The set of equilibria is a submanifold $\mathcal{M}_{eq}$ of the configuration space.
Points in $\mathcal{M}_{eq}$ can be labeled by the corresponding values of $\mu_i$ which 
are called control parameters. Usually all that we know is the explicit expression of the 
equilibrium Massieu function
\begin{equation}
 Z_{eq} = Z (\mu^i),
\end{equation}
which is the integral of Eq.~(\ref{eq:dS}).
The explicit expressions for the conjugate variables at equilibrium are calculated as
\begin{equation}
 \beta_i (\mu^i) = \frac{\partial Z_{eq}}{\partial \mu^i}.
\end{equation}
The entropy maximum postulate is a statement about the behaviour of the entropy function along 
off-equilibrium curves in $\mathcal{M}$, not along equilibrium sequences $\mathcal{M}_{eq}$.
Therefore we need an expression for an extended Massieu function $\hat{Z} = \hat{Z}(X^\rho, \mu^i)$ 
where $X^\rho$ denote a set of off-equilibrium variables.
The equillibrium configurations occur at the values $X^\rho_{eq} = X^\rho_{eq}(\mu^i)$ which 
are solutions of
\begin{equation}
 \frac{\partial \hat{Z}}{\partial X^\rho} = 0.
\end{equation}
A stable equilibrium takes place at the point $X^\rho_{eq}$ if and only if it is a local maximum 
of $\hat{Z}$ at fixed $\mu^i$. Therefore stable solutions have a matrix
\begin{equation}
 \left. \frac{\partial^2 \hat{Z}(X^\tau, \mu^i)}{\partial X^\rho \partial X^\sigma}\right|_{X^\tau = X^\tau_{eq}(\mu^i)}
 \label{eq:matrix}
\end{equation}
with a negative spectrum of eigenvalues.
A change of stability takes place when one of these eigenvalues become zero and changes sign.
It has been shown that we can obtain information about the sign of an eigenvalue near a point 
where it vanishes without computing the spectrum of the matrix (\ref{eq:matrix}).
What we have to do is to plot a conjugate variable 
$\beta_a (\mu^a) = \frac{\partial Z_{eq}}{\partial \mu^a}$ against a control parameter 
$\mu^a$ along an equilibrium sequence for some fixed $a$.
When a change of stability occurs, called a turning point, the plot of the stability 
curve $\beta_a (\mu^a)$ has a vertical tangent. 
Figure~\ref{fig:turning} shows an 
example of stability curves with a turning point $A$.  
The branch with negative slope near this point is always unstable, since one can
prove that at least one eigenvalue of the matrix (\ref{eq:matrix}) is positive.
The branch with positive slope near $A$ is more stable.
If the spectrum of zero modes is nondegenerate, only one of the eigenvalues changes its 
sign at the turning point. Therefore the positive slope branch has one unstable mode less 
than the negative slope branch. Note that what we can see with this method is the existence 
of instabilities because there can be positive eigenvalues which do not change sign at the 
turning point. If there is a point in the equilibrium sequence which is fully stable, all 
equilibria in the sequence are fully stable until the first turning point is reached.

\subsection{Pure black holes}
The case of a black hole isolated from its surroundings can be described in the microcanonical 
emsemble. The Massieu function is the R\'enyi entropy (\ref{eq:Renyi_entropy})
and the control parameters are $M$ and $J$. The conjugate variables are $\beta$ and $-\alpha$, 
where $\beta$ is the inverse of the R\'enyi temperature (\ref{eq:Renyi_temperature}) and 
$\alpha = \frac{\Omega}{T_R}$. To study the stability of the black hole we need to plot 
the stability curves $\beta(M)$ at constant $J$ and $-\alpha(J)$ at constant $M$.

\begin{figure}
  \noindent\hfil\includegraphics[scale=0.7,angle=0]{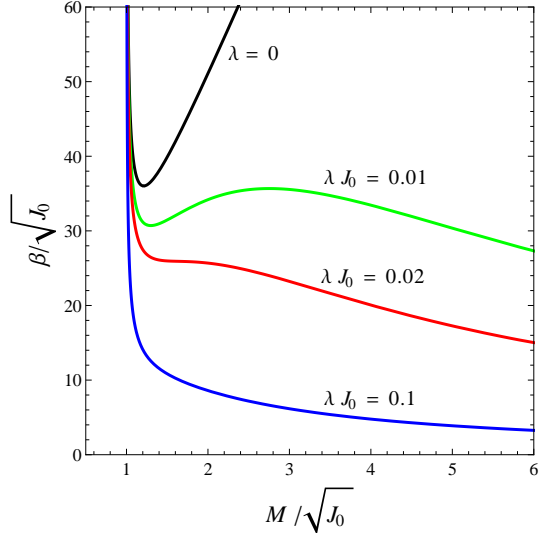}
  \caption{Curves of the conjugate variable $\beta(M)$ at fixed $J$ in the microcanonical 
  treatment. The $\lambda = 0$ (black) curve represents the stability curve of the black 
  hole in the standard thermodynamic approach, while the $\lambda J_0 = 0.01$ (green), 
  $\lambda J_0 = 0.02$ (red) and $\lambda J_0 =0.1$ (blue) curves are the stability curves 
  within the R\'enyi approach. No vertical tangent occurs in either case. By rotating the 
  figure clockwise with $\frac{\pi}{2}$, the stability curves of the canonical treatment 
  can be obtained, i.e.~$-M(\beta)$ at fixed $J$. In this case, the $\lambda J_0 = 0.01$ 
  (green) curve has two vertical tangents denoting the loss and the recovery of stability. 
  In this scenario, up to three black holes with different mass-energy parameters can coexist 
  at a given temperature.}
 \label{fig:micro_bM}
\end{figure}

When $J=J_0$ is a constant, the variables $\beta/\sqrt{J_0}$ and $M/\sqrt{J_0}$ can 
be written as functions of the normalized parameter $h$ with constant $\lambda J_0$.
On Fig.~\ref{fig:micro_bM} we plotted the stability curves $\beta(M)$ for different 
values of $\lambda J_0$. For comparison, the stability curve of the standard Kerr black 
hole ($\lambda = 0$) is also plotted on the figure. It can be seen that no curves
of the Kerr-R\'enyi case have a vertical tangent (similar to the standard 
Kerr result analyzed by Kaburaki, Okamoto and Katz in \cite{Kaburaki:1993ah}),
and therefore there is no stability change for any $M$.
Like in the $\lambda = 0$ case, the $\lambda>0$ curves also diverge asymptotically 
as we consider the extreme black hole limit $h\rightarrow 1$. 
In the large $M$ limit the black holes approach the static Schwarzschild solution 
($h=0$). It can also be seen that the $\lambda > 0$ stability curves are similar to 
the Schwarzschild-R\'enyi case in the large $M$ region (see Fig.~3 of \cite{Czinner:2015eyk}).

For the standard case, it has been shown that isolated Kerr holes are thermodynamically 
stable with respect to axisymmetric perturbations. Also, the isolated Schwarzschild 
black holes have been found to be stable against spherically symmetric perturbations 
in the R\'enyi approach. Based on these results, since no turning point occurs on the
stability curves in between these two extrema, we can conclude that isolated Kerr black 
holes are thermodynamically stable against axisymmetric perturbations in the R\'enyi 
approach as well.

\begin{figure}
  \noindent\hfil\includegraphics[scale=0.7,angle=0]{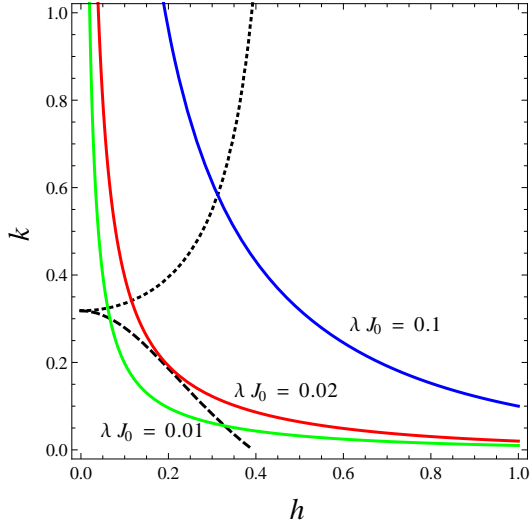}
  \caption{Plots of $J = \mbox{const.}$ curves for $\lambda J_0 = 0.01$ (green), 
  $\lambda J_0 = 0.02$ (red) and $\lambda J_0 =0.1$ (blue) parameter values on 
  the $(h,k)$ space.}
 \label{fig:hc_J0}
\end{figure}

By looking at Fig.~\ref{fig:micro_bM}, we can also see that there are two points where the 
tangent of the stability curves with smaller $\lambda$ (or small angular momentum $J_0$) 
becomes horizontal. These correspond to the points where the heat capacity at constant $J$ 
changes its sign through an infinite discontinuity, similar to the Davies point {\cite{Davies:1978mf}} 
of the standard Kerr black hole case. On Fig.~\ref{fig:hc_J0} we plotted 
the lines of constant $J$ in the normalized parameter space of $(h,k)$. Here the 
$\lambda J_0 = 0.01$  line crosses the line where $C_{JR}$ diverges. Along this line the heat 
capacity $C_{JR}$ changes its sign two times on the way from the $(h=0)$ Schwarzschild limit 
to $(h = 1)$ extremality.

\begin{figure}
  \noindent\hfil\includegraphics[scale=0.7,angle=0]{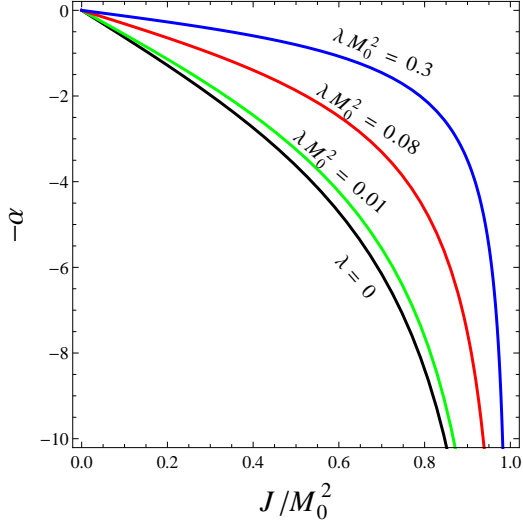}
  \caption{Curves of the conjugate variable $-\alpha(J)$ at fixed $M$. The $\lambda = 0$ (black) curve
  represents the stability curve of a Kerr black hole in the standard approach within the microcanonical 
  treatment. The $\lambda M_0^2 = 0.01$ (green), $\lambda M_0^2 = 0.08$ (red) and $\lambda M_0^2 =0.3$ 
  (blue) curves are the stability curves of the R\'enyi approach. No vertical tangents occurs.}
 \label{fig:micro_maJ}
\end{figure}

\begin{figure}
  \noindent\hfil\includegraphics[scale=0.7,angle=0]{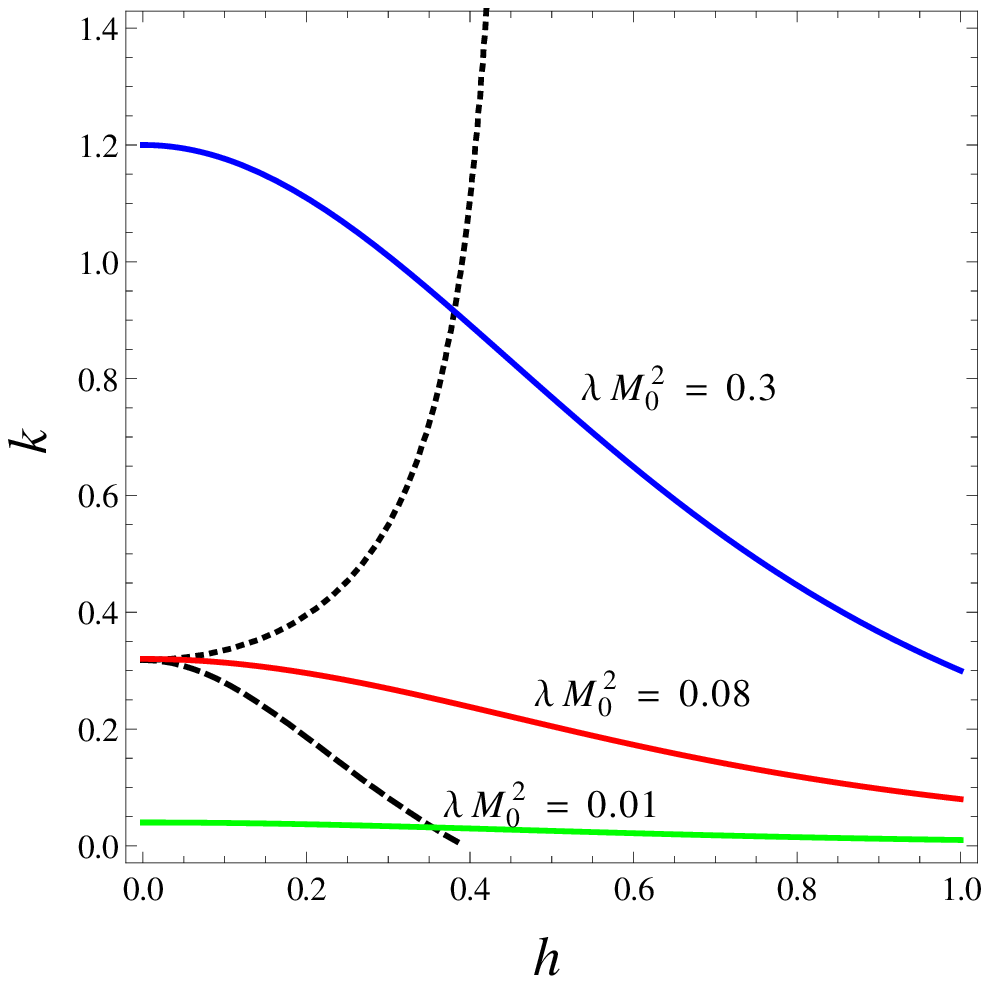}
  \caption{Plots of $M = \mbox{const.}$ curves for $\lambda M_0^2 = 0.01$ (green), 
  $\lambda M_0^2 = 0.08$ (red) and $\lambda M_0^2 =0.3$ (blue) parameter values on 
  the $(h,k)$ space.}
 \label{fig:hc_M0}
\end{figure}

When $M=M_0$ is a constant, $\alpha$ and $J/M_0^2$ become functions of $h$ with a constant 
$\lambda M_0^2$. On Fig.~\ref{fig:micro_maJ}, we plotted the $-\alpha(J)$ stability curves 
for different values of $\lambda M_0^2$. There is no vertical tangent and hence no stability 
change occurs at any point for any $\lambda$. The $M=const.$ lines in the parameter space 
of $(h,k)$ are plotted on Fig.~\ref {fig:hc_M0}.

\subsection{Black holes in a heat bath}
\label{sec:heat_bath}

Let us now consider the black hole in the canonical approach. The canonical ensemble 
describes the system of a black hole in equilibrium with an infinite reservoir of 
thermal radiation at constant temperature. The Massieu function in this case is
\begin{equation}
 \Psi(\beta,J) = S_R - \beta M = -\beta F,
\end{equation}
where $F = M - T_R S_R$ is the Helmholtz free energy. The conjugate variables of the control 
parameters are $-M$ and $-\alpha$. To study the stability of the black hole we need to plot 
the stability curves $-M(\beta)$ at constant $J$ and $-\alpha(J)$ at constant $\beta$.

The stability curves of $-M(\beta)$ at constant $J$ are simply the $\frac{\pi}{2}$ clockwise 
rotated versions of Fig. \ref{fig:micro_bM}. We can see that there is a vertical tangent along 
the stability curve in the standard Boltzmann treatment ($\lambda = 0$).
The heat capacity $C_J$ diverges at this turning point where $h=h_c$.
The $0<h<h_c$ branch of this curve is less stable than the $h>h_c$ branch.
As Kaburaki, Okamato and Katz have shown \cite{Kaburaki:1993ah}, one can conclude from this result that 
{since Schwarzschild black holes ($h=0$) are unstable in an infinite bath, so are the
slowly rotating holes until the $h_c$ turning point is reached. Rapidly rotating Kerr black holes, 
on the other hand, can become stable if the slowly rotating (unstable) 
holes have only one negative eigenmode, which changes sign at $h_c$.}

For the parametrized R\'enyi case, the behavior of the stability curves changes depending 
on the value of $\lambda J_0$. We can see that there are two turning points on the  
 $\lambda J_0 = 0.01$ stability curve when we rotate the plots of Fig.~\ref{fig:micro_bM} with $\pi/2$ 
clockwise. These turning points disappear when the value of $\lambda J_0$ is larger than a critical 
value. As a consequence, stability change occurs only when the parameter $\lambda$ and/or $J_0$
is sufficiently small. In this case, there are three phases of black holes; small, intermediate 
unstable and large black holes.
The stability property of a rapidly rotating, small black hole is the same as
of a slowly rotating, large black hole, which is expected to be stable from continuity requirements
to the static solution in the R\'enyi approach {\cite{Czinner:2015eyk}}.

\begin{figure}
  \noindent\hfil\includegraphics[scale=0.7,angle=0]{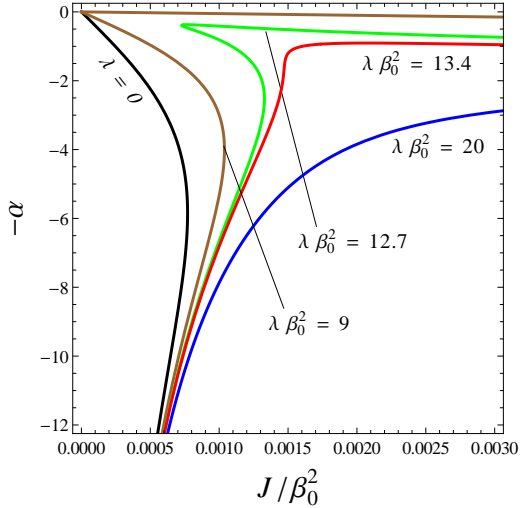}
  \caption{Curves of the conjugate variable $-\alpha(J)$ at fixed $\beta$ in the canonical approach. 
  The $\lambda = 0$ (black) curve describes the standard thermodynamic approach. The  
  $\lambda \beta_0^2 = 9$ (brown), $\lambda \beta_0^2 = 12.7$ (green), $\lambda \beta_0^2 = 13.4$ 
  (red) and $\lambda \beta_0^2 = 20$ (blue) curves are the stability curves of the R\'enyi model. 
  The $\lambda \beta_0^2 = 12.7$ (green) curve has two turning points, while the stability 
  curve of $\lambda \beta_0^2 = 9$ (brown) exhibits two branches with a single turning point in 
  the lower branch.}
 \label{fig:cano_maJ}
\end{figure}

\begin{figure}
  \noindent\hfil\includegraphics[scale=0.7,angle=0]{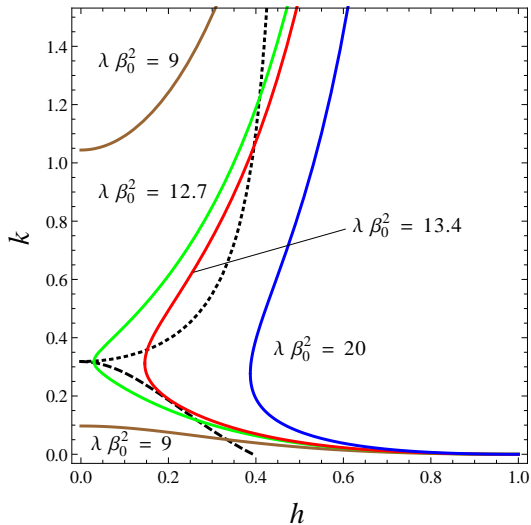}
  \caption{Plots of $\beta = \mbox{const.}$ curves on the $(h,k)$ space for $\lambda \beta_0^2 = 9$ 
  (brown), $\lambda \beta_0^2 = 12.7$ (green), $\lambda \beta_0^2 = 13.4$ (red) and $\lambda \beta_0^2 = 20$ (blue).}
 \label{fig:hc_beta0}
\end{figure}

The stability curves of $-\alpha(J)$ at constant $\beta$ are plotted on Fig.~\ref{fig:cano_maJ},
while the lines of constant $\beta$ in the parameter space of $(h,k)$ are plotted on Fig.~\ref{fig:hc_beta0}.
One can see again that the stability curves have no vertical tangent when the $\lambda$ parameter 
is sufficiently large, similar to the case of $\beta(M)$ with constant $J$. Vertical tangents to 
the stability curves appear however when $\lambda$ and/or $\beta_0$ are smaller than some critical 
value. When a curve has two vertical tangents, there are two turning points where the heat capacity 
$C_{JR}$ diverges and changes its sign. The black hole changes its stability there in the order from 
{a less unstable state to unstable state and back to a less unstable state again, 
where the less unstable states may even be stable but not guaranteed.}

{When $\lambda \beta_0^2$ is less than
some critical value, there are two branches of the stability curves $-\alpha(J)$ at constant 
$\beta_0$. This result is consistent with the fact that in the Schwarzschild-R\'enyi case there 
are two black holes with the same $\beta$, as it can be seen on Fig.~1 in \cite{Czinner:2015eyk}. 
The lower branch of the $\lambda \beta_0^2=9$ curve terminates at the $h = 0$ small, static black 
hole limit. The behavior of this branch is similar to the curve of the $\lambda = 0$ 
Kerr-Boltzmann case. 
According to these results, we can conclude that small, static or slowly rotating 
black holes in the R\'enyi approach are unstable in a heat bath, but fast rotation 
can stabilize them in a similar way as it is done in the Kerr-Boltzmann case, which has 
been shown by Kaburaki, Okamoto and Katz \cite{Kaburaki:1993ah}. 
The upper branch of the $\lambda \beta_0^2=9$ stability curve 
belongs to larger mass black holes and terminates at the large, static black hole limit when 
$h \rightarrow 0$. There is no vertical tangent in this branch so the corresponding rotating 
black holes have the same stability property as the large, static black holes in a heat bath,
i.e.~they are stable.}

\subsection{Phase Transitions}
\label{sec:phase}
We have shown in our earlier paper \cite{Czinner:2015eyk} that a Hawking--Page phase 
transition can be observed for static black holes in the R\'enyi approach. Previously, it 
had also been shown \cite{Caldarelli:1999xj,Tsai:2011gv,Altamirano:2014tva} that Kerr-AdS 
black holes exhibit a first order small black hole/large black hole (SBH/LBH) phase transition 
in the canonical ensemble. In this subsection we will study the question of possible phase 
transitions of Kerr black holes in the R\'enyi model.

\begin{figure}
  \noindent\hfil\includegraphics[scale=0.7,angle=0]{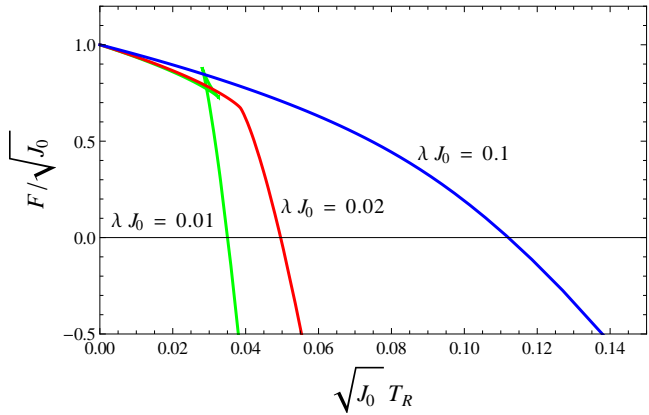}
  \caption{Free energy of a Kerr black hole in the R\'enyi model against the temperature 
  for various angular momenta $J_0$, $\lambda J_0 = 0.01$ (green), $\lambda J_0 = 0.02$ (red) 
  and $\lambda J_0 = 0.1$ (blue). Characterristic swallowtail behaviour is observed for 
  $\lambda J_0 = 0.01$ (green), which corresponds to a SBH/LBH phase transition.}
 \label{fig:FT_Jconst}
\end{figure}

The behavior of the free energy function $F = M - T_R S_R$ for $\lambda J_0 =$ 0.01, 0.02 
and 0.1 at constant $J$ is displayed on Fig.~\ref{fig:FT_Jconst}. We can see that all curves 
cross the horizontal axis. Small black holes with lower temperature possess positive free energy, 
while larger black holes with higher temperature possess negative free energy. One can, therefore, 
expect a Hawking--Page transition between the thermal gas phase with angular momentum, 
and the large black hole state, which would be locally stable according to our analysis in 
Sec.~\ref{sec:heat_bath}. It is generally assumed that $F\approx 0$ for a thermal gas, so the 
phase transition occurs around the temperature where the free energy of the 
black hole becomes zero.

\begin{figure}
  \noindent\hfil\includegraphics[scale=0.7,angle=0]{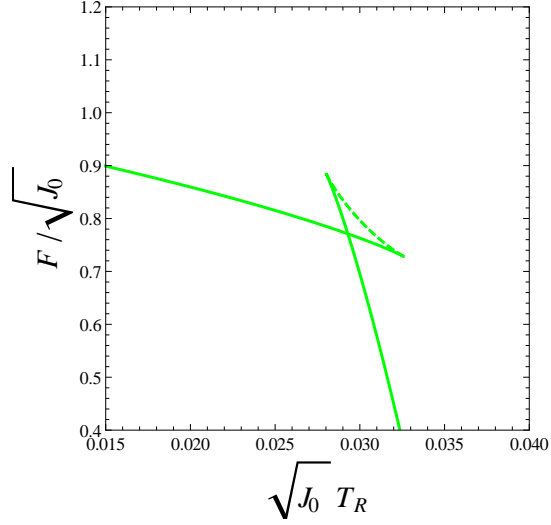}
  \caption{Close up figure of the free energy of a Kerr black hole in the R\'enyi 
  model for $\lambda J_0 = 0.01$ on Fig \ref{fig:FT_Jconst}. The intermediate, unstable 
  branch is displayed with a dashed line.}
 \label{fig:FT_swallow}
\end{figure}

An SBH/LBH phase transition can also be observed for Kerr black holes in the R\'enyi model
when we enlarge the $\lambda J_0 =$ 0.01 curve of Fig.~\ref{fig:FT_Jconst} on 
Fig~\ref{fig:FT_swallow}. The swallowtail behavior of the free energy function is a 
typical sign of a first order transition between the SBH and LBH phases. There
are three branches on the picture: small, lower temperature holes; large, higher temperature 
holes; and also intermediate, unstable black holes. There is a coexistence point of small and 
large black holes where the SBH/LBH transition occurs. The mass and entropy functions are 
discontinuous at this point which indicates that the phase transition is a first order kind.
By increasing $\lambda$, the swallowtail behavior disappears, as it can be seen on the 
$\lambda J_0 =$ 0.02 curve on Fig.~\ref{fig:FT_Jconst}. This suggests the existence 
of a critical point where the phase transition becomes second order.

\section{Kerr-AdS black holes}
\label{sec:Kerr-AdS}

\begin{figure}
  \noindent\hfil\includegraphics[scale=0.7,angle=0]{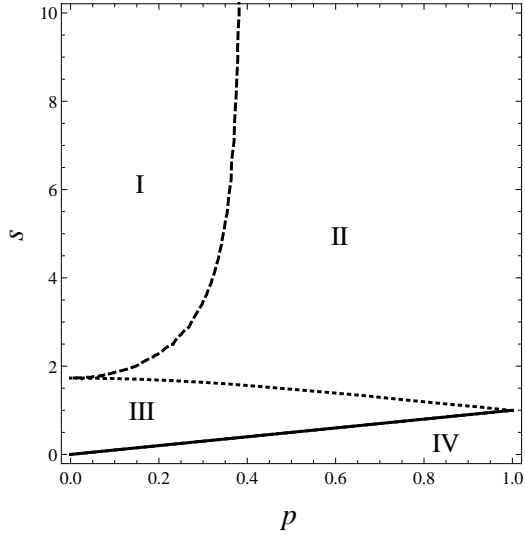}
  \caption{Phase diagram of Kerr-AdS black holes in the standard model. 
  On the dashed curve $C_J$, while on the dotted curve $C_\Omega$ diverges. 
  In region I, $C_{\Omega}<0$ and $C_{J} <0$, in region II, $C_{\Omega}<0$
  and $C_{J} >0$, and in region III, $C_{\Omega}>0$ and $C_{J} >0$. In region 
  IV, there is no physical solution because $|a| > l$.}
 \label{fig:heat_capacity_AdS}
\end{figure}

In order to compare the obtained stability results of the Kerr-R\'enyi model to the 
Kerr-AdS-Boltzamann case in the Poincar\'e approach (analogous to the Schwarzschild 
problem), in this section we present the Poincar\'e stability analysis of the 
Kerr-AdS-Boltzamann case as well. The Kerr-AdS black hole metric is described by
\begin{eqnarray}
 ds^2 & = & - \frac{\Delta_r}{\rho^2} \left( dt - \frac{a \sin^2 \theta}{\Xi} d \phi \right)^2 
 + \frac{\rho^2}{\Delta_r} dr^2 + \frac{\rho^2}{\Delta_\theta} d \theta^2 \nonumber \\
 & & + \frac{\Delta_\theta \sin^2 \theta}{\rho^2} \left( a dt - \frac{r^2 + a^2}{\Xi} d\phi \right)^2,
\end{eqnarray}
where
\[
\Delta_r = (r^2 + a^2) \left( 1 + \frac{r^2}{l^2} \right) - 2 m r, ~~~ \Delta_\theta
= 1 - \frac{a^2 \cos^2 \theta}{l^2}, ~~~
\]
\[
\rho^2 = r^2 + a^2 \cos^2 \theta, ~~~ \Xi = 1 - \frac{a^2}{l^2}.
\]
The thermodynamic quantities are written in terms of  $a$, $l$ and the horizon radius $r_{+}$, 
which is obtained by solving $\Delta = 0$. The Hawking temperature of the horizon is given by
\begin{equation}
 T = \frac{1}{2 \pi r_{+}} \left( \frac{(a^2 +3 r_{+}^2)(r_{+}^2/l^2+1)}{2(a^2 + r_{+}^2)} -1\right),
\end{equation}
while the Bekenstein-Hawking entropy of the black hole is
\begin{equation}
 S =\pi \frac{a^2 + r_{+}^2}{1 -a^2/l^2}.
\end{equation}
The angular momentum of a Kerr-AdS black hole is
\begin{equation}
 J = \frac{(r_{+}^2 + a^2)(1 + r_{+}^2/l^2)}{2 r_{+}} \frac{a}{(1 - a^2/l^2)^2},
\end{equation}
the angular velocity of the horizon is
\begin{equation}
 \Omega = \frac{a}{l^2} \frac{r_{+}^2 + l^2}{r_{+}^2 + a^2},
\end{equation}
and the mass-energy parameter of the black hole can be re-expressed as
\begin{equation}
 M = \frac{(r_{+}^2 + a^2)(1 + r_{+}^2/l^2)}{2 r_{+}} \frac{1}{(1 - a^2/l^2)^2}.
\end{equation}
The heat capacity at constant angular velocity can be computed as
\begin{equation}
 C_\Omega = \frac{2 \pi l^2 r_{+}^2( 3 r_{+}^4 +( a^2 + l^2) r_{+}^2 
 - a^2 l^2)}{(l^2 - a^2)(3 r_{+}^4 -(a^2 + l^2) r_{+}^2 - a^2 l^2)},
\end{equation}
and the heat capacity at constant angular momentum takes the form
\begin{equation}
 C_J = \frac{2 \pi l^4  \left(a^2+r_{+}^2\right)^2 \left(-a^2 l^2+\left(a^2+l^2\right) r_{+}^2
 +3 r_{+}^4\right)}{(l^2 -a^2) X},
\end{equation}
where
\begin{eqnarray*}
 X & = -l^4 r_{+}^4+3 l^2 r_{+}^6+a^6 \left(l^2+r_{+}^2\right) +a^4 \left(3 l^4+13 l^2 r_{+}^2+6 r_{+}^4\right) \\
   & \quad +a^2 \left(6 l^4 r_{+}^2+23 l^2 r_{+}^4+9 r_{+}^6\right).
\end{eqnarray*}
Here we introduced the normalized parameters
\begin{equation}
 p \equiv \frac{|a|}{r_{+}}, ~\mbox{and}~~ s \equiv \frac{l}{r_{+}}.
\end{equation}
The heat capacities $C_\Omega$ and $C_J$ change their signs depending on the values of $p$ and $s$.
The parameter space of $(p,s)$ can be divided into 4 regions depending on the signs of $C_\Omega$ and 
$C_J$ as shown on Fig.~\ref{fig:heat_capacity_AdS}.

Similarly to the Kerr-R\'enyi case in Sec.~\ref{sec:Stability}, the thermodynamic stability problem 
of Kerr-AdS black holes can also be analyzed by the Poincar\'e turning point method. The stability 
curves of the two systems are qualitatively similar. For the study of the Kerr-AdS black hole 
problem we will use the following normalized variables,
\begin{equation}
 \tilde{\beta} = \frac{\beta}{l},~~ 
 \tilde{J} = \frac{J}{l^2}, ~~
 \tilde{M} = \frac{M}{l}.
\end{equation}

\begin{figure}
  \noindent\hfil\includegraphics[scale=0.7,angle=0]{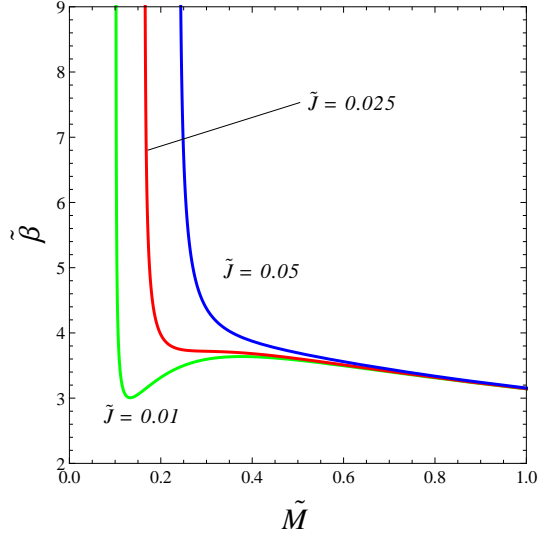}
  \caption{Stability curves of $\beta(M)$ at fixed $J$ for Kerr-AdS black holes in the microcanonical 
  treatment. The curves of  $\tilde{J} = 0.01$ (green), $\tilde{J} = 0.025$ (red), and $\tilde{J} =0.05$ 
  (blue) are plotted. No vertical tangent occurs in either case. The figure rotated by $\frac{\pi}{2}$ 
  clockwise represents the stability curves of $-M(\beta)$ at fixed $J$ for the canonical ensemble, 
  in which case the $\tilde{J} = 0.01$ (green) curve has two vertical tangents.}
 \label{fig:micro_bM_ads}
\end{figure}

\begin{figure}
  \noindent\hfil\includegraphics[scale=0.7,angle=0]{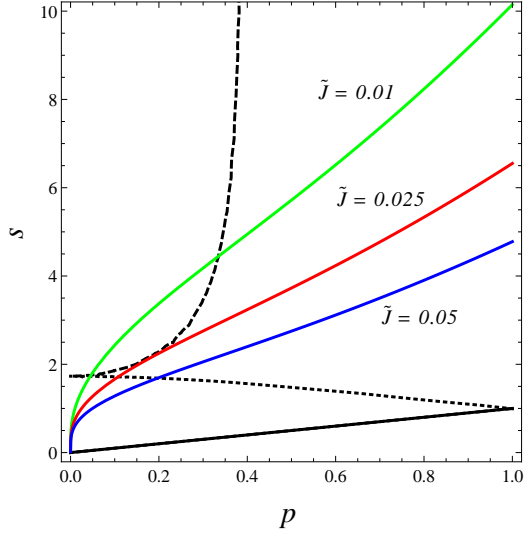}
  \caption{Plots of $J = \mbox{const.}$ curves for $\tilde{J} = 0.01$ (green), $\tilde{J} = 0.025$ 
  (red) and $\tilde{J} =0.05$ (blue) on the $(p,s)$ space.}
 \label{fig:hc_J_ads}
\end{figure}

\begin{figure}
  \noindent\hfil\includegraphics[scale=0.7,angle=0]{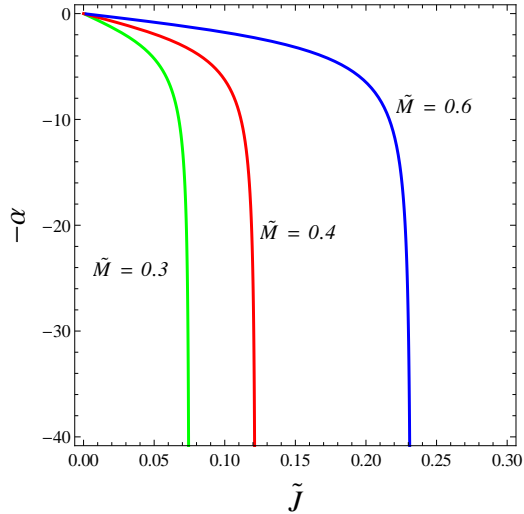}
  \caption{Curves of the conjugate variable $-\alpha(J)$ at fixed $\beta$ for Kerr-AdS black holes 
  in the canonical treatment. The stability curves of $\tilde{M} = 0.3$ (green), $\tilde{M} = 0.4$ 
  (red) and $\tilde{M} = 0.6$ (blue) are plotted. There are no turning points on the diagram.}
 \label{fig:micro_maJ_ads}
\end{figure}

\begin{figure}
  \noindent\hfil\includegraphics[scale=0.7,angle=0]{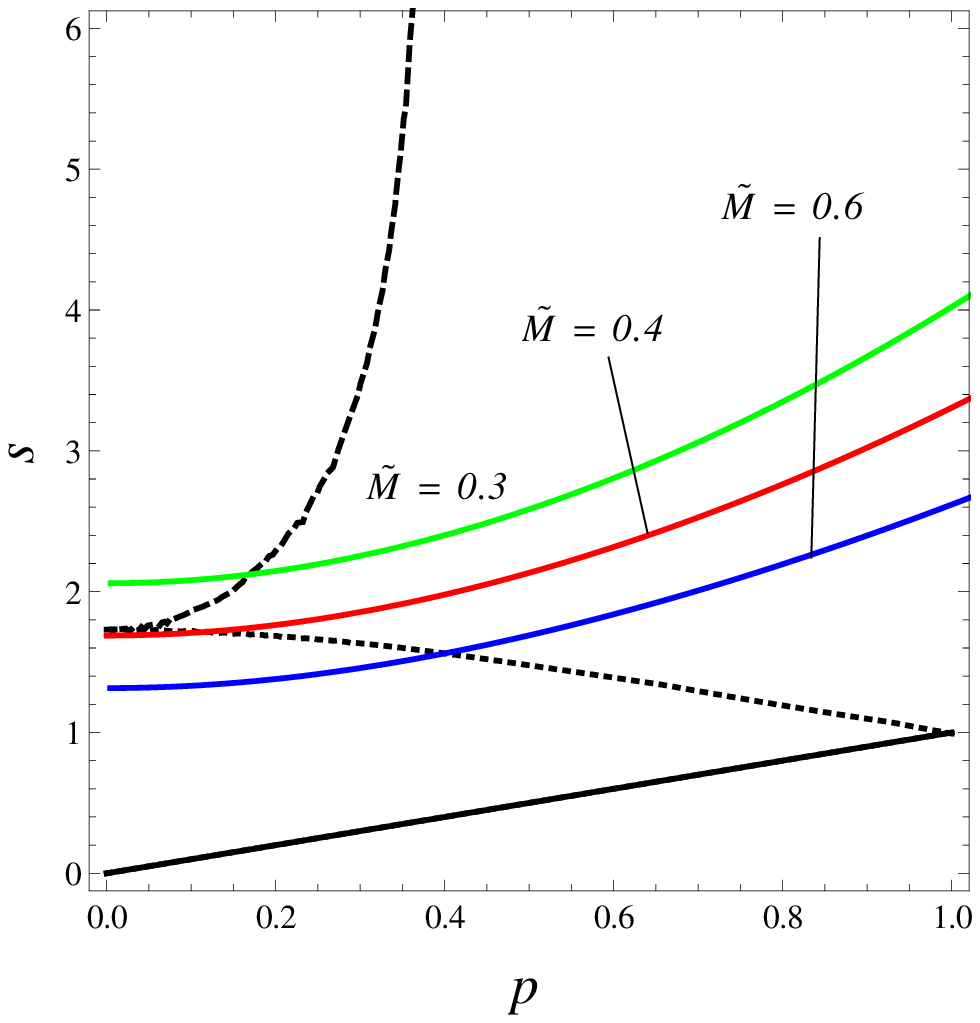}
  \caption{Plots of $M = \mbox{const.}$ curves for $\tilde{M} = 0.3$ (green), $\tilde{M} = 0.4$ (red) 
  and $\tilde{M} = 0.6$ (blue) on the $(p,s)$ plane.}
 \label{fig:hc_M0_ads}
\end{figure}

First we consider the microcanonical ensemble. The stability curves $\beta(M)$ at constant $J$ are plotted 
on Fig.~\ref{fig:micro_bM_ads}, while the curves of constant $J$ in the parameter space of $(p,s)$ are 
plotted on Fig.~\ref{fig:hc_J_ads}. Just like in the Kerr-R\'enyi case, we can see that there is no 
turning point of stability. The stability curves $-\alpha(J)$ at constant $M$ and the curves of constant 
$M$ in the $(p,s)$ space are depicted on Fig.~\ref{fig:micro_maJ_ads} and Fig.~\ref{fig:hc_M0_ads} respectively.
The behavior of the stability curves is almost identical to the one of the Kerr-R\'enyi case.

\begin{figure}[!htb]
\noindent\hfil\includegraphics[scale=0.7,angle=0]{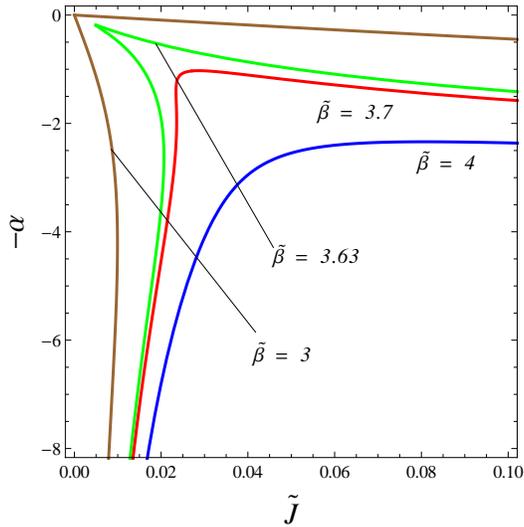}
  \caption{Curves of the conjugate variable $-\alpha(J)$ at fixed $\beta$ for Kerr-AdS black holes in 
  the canonical ensemble. The curves of $\tilde{\beta} = 3$ (brown), $\tilde{\beta} = 3.63$ (green), 
  $\tilde{\beta} = 3.7$ (red) and 
  $\tilde{\beta} = 4$ (blue) are plotted. The $\tilde{\beta} = 3.63$ (green) curve has two 
  turning points, while the $\tilde{\beta} = 3$ (brown) curve has two branches and the lower 
  branch has a turning point.}
 \label{fig:cano_maJ_ads}
\end{figure}

\begin{figure}[!htb]
\noindent\hfil\includegraphics[scale=0.7,angle=0]{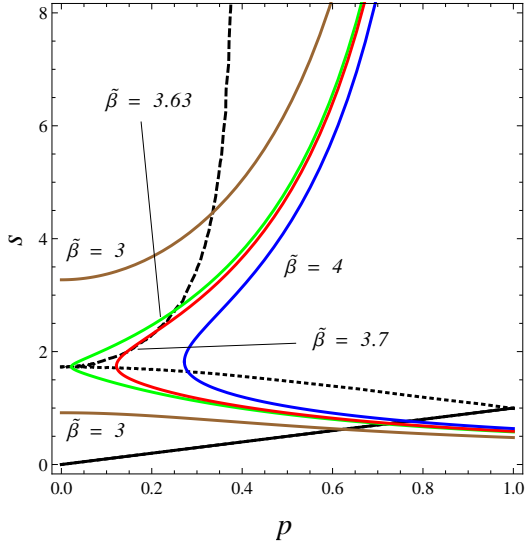}
\caption{Plots of $\beta = \mbox{const.}$ curves on the $(p,s)$ space for  $\tilde{\beta} = 3$ 
(brown), $\tilde{\beta} = 3.63$ (green), $\tilde{\beta} = 3.7$ (red) and $\tilde{\beta} = 4$ (blue).}
\label{fig:hc_beta0_ads}
\end{figure}

For the canonical system, the stability curves of $-M(\beta)$ at constant $J$ can be seen on 
Fig.~\ref{fig:micro_bM_ads} if we rotate it by $\frac{\pi}{2}$ clockwise.
The figure shows the existence of a critical temperature, above which the Kerr-AdS black holes 
allow a first order SBH/LBH phase transition in the canonical ensemble.
On Fig.~\ref{fig:cano_maJ_ads} we plotted the stability curves of $-\alpha(J)$ at constant $\beta$.
The curves of constant $\beta$ on the $(p,s)$ space are plotted on Fig.~\ref{fig:hc_beta0_ads}.
There are no turning points on the lower temperature (larger $\beta$) curves,
but higher temperature curves exhibit turning points. 
Therefore, a stability change of Kerr-AdS black holes occurs only when the temperature is higher 
than a certain critical value. Black holes with slightly higher temperature than the critical one
have an unstable branch between two, more stable branches. There is another critical temperature 
above which a cusp appears on the stability curve at $(\tilde{J},-\alpha) = (0,0)$, where the 
Kerr-AdS black hole reduces to the Schwarzschild-AdS case. A vertical tangent occurs in the 
small black hole branch only, and no vertical tangent exists in the large black hole branch.
From this result we can conclude that small and slowly rotating Kerr-AdS black holes are unstable 
in the canonical ensemble.

As it can be clearly seen from the analysis above, the thermodynamic properties of the Kerr-R\'enyi and 
the Kerr-AdS-Boltzmann models are very similar. In the static case, we have obtained a simple relation 
between the entropy parameter $\lambda$ and the AdS curvature parameter $l$ for black holes with 
identical horizon temperatures \cite{Biro:2013cra}. By assuming the same condition for stationary 
black holes augmented with the assumption of identical horizon angular velocity, we can derive  
analogous relations between the $(h,k)$ and $(p,s)$ parameters for rotating black holes by solving 
the following equations 
\begin{equation}
 \hat{T}_R (h,k) = \hat{T}_{AdS} (p,s), ~~~ \hat{\Omega}_R (h,k) = \hat{\Omega}_{AdS} (p,s),
\end{equation}
where we normalized the quantities by the horizon radius $r_{+}$ as
\begin{equation}
 \hat{T} = T r_{+}, ~~~ \hat{\Omega} = \Omega r_{+}.
\end{equation}
As a result, a quantitative analogy between the Kerr-R\'enyi and Kerr-AdS-Boltzmann 
pictures of black hole thermodynamics can be given by the parameter equations 
\begin{eqnarray}
 p & = & \frac{3+h^2+k \pi -h^4 k \pi -Y}{2 h}, \\
 s & = &  \sqrt{\frac{3(1+h^2)}{3-h^2+2 k \pi -2 h^4 k \pi 
 -Y}},
\end{eqnarray}
where
\[
 Y = \sqrt{\left(3-h^2\right)^2+2 \left(3+h^2-3 h^4-h^6\right) k \pi 
 +\left(1-h^4\right)^2 k^2 \pi ^2}.
\]
These equations provide a very interesting correspondence between the two approaches.

\section{Summary and Conclusions}
\label{sec:summary}

In this paper we investigated the thermodynamic and stability properties of Kerr black holes 
described by the parametric, equilibrium- and zeroth law compatible R\'enyi entropy function.
The corresponding problem of static Schwarzschild
black holes has been analyzed in \cite{Biro:2013cra, Czinner:2015eyk}, where interesting
similarities have been found to the picture of standard black hole thermodynamics in asymptotically
AdS space. In particular, a stability change and a Hawking--Page transition have been identified,
which motivated us to extend our investigations to the present (3+1)-dimensional, rotating problem 
as well.  

The novel results of this work are the following. We derived the temperature and heat capacities 
of a Kerr black hole in the R\'enyi approach, and found that the global maximum of the temperature-energy 
curve at a fixed angular momentum in the standard description becomes only a local maximum in 
the R\'enyi model. In the thermodynamic stability analysis we investigated both the microcanonical 
and the canonical ensembles. We have plotted the stability curves of the Boltzmann-Gibbs and R\'enyi 
entropy models, and showed that no stability change occurs for isolated black holes in either case. 
From this result, we concluded that, similarly to the standard Boltzmann case, isolated Kerr black 
holes are thermodynamically stable with respect to axisymmetric perturbations in the R\'enyi approach. 

In case when the black holes are surrounded by a bath of thermal radiation in the canonical 
picture, we found that, in contrast to the standard Boltzmann case, slowly rotating Kerr black 
holes can be in stable equilibrium with thermal radiation at a fixed temperature if the number of 
negative eigenmodes of the stability matrix is one. We showed that fast rotating black holes have 
similar stability properties to slowly rotating ones, and there may also exist intermediate size, 
unstable black holes. We also analyzed the question of possible phase transitions in the canonical 
picture, and found that, in addition to a Hawking--Page transition, a first order small black 
hole/large black hole phase transition occurs in a very similar fashion as in AdS space.
These findings indicate that there is a similarity between the Kerr-R\'enyi and Kerr-AdS-Boltzamann 
models, analogous to the one that we found in the static case. Based on this result we also 
investigated the Poincar\'e stability curves of Kerr-AdS black holes in the standard Boltzmann 
picture, and confirmed this similarity by obtaining simple algebraic relations between the parameters 
of the two approaches with identical surface temperature and angular velocity. 

The above results may be relevant in many aspects of black hole physics. Our main motivation 
in the first place was to consider a statistical model to the nonextensive and nonlocal nature 
of black hole thermodynamics, where we do not assume \emph{a priori} that the classical, 
additive Boltzamann statistics can describe this strongly gravitating system.
The R\'enyi form of the black hole entropy includes a parameter $\lambda$, which seems to be 
a good candidate to incorporate the effects of the long-range type behavior of the gravitational
field, while also being additive and satisfying both the equilibrium 
compatibility and the zeroth law's requirements. A specific model on how to compute the 
$\lambda$ parameter value for the black hole problem is yet to be developed, but a similar 
approach has been considered in \cite{CzM} to describe the mutual information
between spatially separated, compact domains of an inhomogeneous universe that are entangled
via the gravitational field equations. In that work,
as an effective model, the $\lambda$ parameter of the Tsallis/R\'enyi relative entropy 
has been defined in a geometric way in order to describe the causal connection between 
the domain and its surroundings during the cosmic evolution. Since black holes are 
essentially the final states of cosmic structure formation, one can expect that the two 
directions might be connected somehow in the nonlinear regime of matter collapse.

As a different direction, it is also interesting to mention that by considering the Boltzamann
picture in the standard description, the Bekenstein-Hawking entropy has a nontrivial nonadditive
property which also satisfies Abe's formula. In the case of Schwarzschild black holes this nonadditivity 
reads as $H_{\lambda}(S)=\sqrt{S}$ and for Kerr black holes $H_{\lambda}(S)=\frac{S}{\sqrt{S-a^2\pi}}$ 
with $\lambda = 0$. The corresponding thermodynamic and stability problems (by also applying the 
formal logarithm method) has been studied in \cite{czinner2015black} and \cite{CzIg}, respectively.

In the present parametric approach however, the most important result is the confirmation of a stability 
change and the Hawking--Page transition of Kerr black holes in the R\'enyi model. As we discussed
in the introduction, this phenomena has many interesting connections with other open
problems in theoretical physics, e.g.~the cosmic nucleation of matter into black holes 
in the early universe, or due to the similarity to the AdS-Boltzmann problem, it may also 
be connected to the AdS/CFT correspondence and related phenomena. A further motivation 
arises from a different possible interpretation of the parametric R\'enyi picture 
originating from finite size reservoir effects in the canonical ensemble. In a recent 
paper \cite{Biro:2012bka}, 
Bir\'o showed that from the requirement of zero mutual information between a finite 
subsystem and a finite reservoir in thermodynamic equilibrium, the Tsallis- and 
R\'enyi entropy formulas arise very naturally. Although we haven't worked out the
details of this approach yet, it provides a nice possible interpretation
of our findings as placing a black hole into a finite heat bath in the canonical
approach instead of an infinite reservoir (which is an idealistic model), and 
require zero mutual information between the black hole and the reservoir in thermal 
equilibrium. In this situation the system is dominated by the bath, and Bir\'o showed 
that the entropy parameter in this case is proportional to the heat capacity of the 
bath as $\lambda=1/C_0$, where instead of the classical infinite approximation, the 
heat capacity of the bath is a large but finite constant $C_0$. This approach has 
been investigated e.g.~for the case when a quark-gluon plasma system is connected to 
a finite heat bath in \cite{Biro:2013qea}.

In conclusion, several interesting consequences can be deduced from the R\'enyi
approach to black hole thermodynamics which is motivated by various 
physical considerations. Parametric corrections to the black hole entropy problem 
also arise from quantum considerations, e.g.~from string theory, loop quantum gravity 
or other semi-classical theories (see e.g.~\cite{Carlip:2014pma} and references
therein), and we expect that other parametric situations are also possible which
might be connected to the parametric R\'enyi description.

\acknowledgments
The research leading to this result was supported by JSPS via an Invitation Fellowship for Research 
in Japan (Long-term) (No.~L14710) and a Grant-in-Aid for Scientific Research (C) (No.~23540319).
V.G.Cz thanks to Funda\c c\~ao para a Ci\^encia e Tecnologia (FCT) Portugal, for financial 
support through Grant~No.~UID/FIS/00099/2013.


\begin{thebibliography}{99}

\bibitem{Hawking:1982dh} S.W.~Hawking and D.~Page, \emph{Thermodynamics of Black Holes in anti-De Sitter Space},
\emph{Commun.~Math.~Phys.}~{\bf 87} (1983) 577.
\bibitem{Maldacena:1997re} J.M.~Maldacena, \emph{The Large N limit of superconformal field theories and supergravity}, 
\emph{Int.~J.~Theor.~Phys.}~{\bf 38} (1999) 1113.
\bibitem{Witten:1998qj} E.~Witten, \emph{Anti-de Sitter space and holography}, 
\emph{Adv.~Theor.~Math.~Phys.}~{\bf 2} (1998) 253.
\bibitem{MMT1} D.~Mateos, R.C.~Myers and R.M.~Thomson, \emph{Holographic Phase Transitions with Fundamental Matter}, 
\emph{Phys.~Rev.~Lett.}~{\bf 97} (2006) 091601.
\bibitem{MMT2} D.~Mateos, R.C.~Myers and R.M.~Thomson, \emph{Thermodynamics of the brane}, 
\emph{JHEP} {\bf 05} (2007) 067.
\bibitem{Bekenstein:1973ur} J.D.~Bekenstein, \emph{Black holes and entropy}, 
\emph{\prd} {\bf 7} (1973) 2333.
\bibitem{Bardeen:1973gs} J.M.~Bardeen et al., \emph{The Four laws of black hole mechanics}, 
\emph{Commun.~Math.~Phys.}~{\bf 31} (1973) 161.
\bibitem{Hawking:1974sw} S.W.~Hawking, \emph{Particle Creation by Black Holes}, 
\emph{Commun.~Math.~Phys.}~{\bf 43} (1975) 199;
\bibitem{Hawking:1976de} S.W.~Hawking, \emph{Black Holes and Thermodynamics},
\emph{\prd} {\bf 13} (1976) 191.
\bibitem{PhysRevE.83.061147} T.S.~Bir\'o and P.~V\'an, \emph{Zeroth law compatibility of nonadditive thermodynamics},
\emph{\pre} {\bf 83} (2011) 061147.
\bibitem{Mackey} M.C.~Mackey, \emph{Time's Arrow: The Origins of Thermodynamic Behaviour},
Springer, New York, (1992).
\bibitem{T2} C.~Tsallis, \emph{Introduction to Non-Extensive Statistical Mechanics: 
Approaching a Complex World}, Springer (2009).
\bibitem{B1}  T.S.~Bir\'o, \emph{Abstract composition rule for relativistic kinetic 
energy in the thermodynamical limit}, \emph{Europhys.~Lett.}~{\bf 84} (2008) 56003.
\bibitem{Landsberg1984} P.T.~Landsberg, \emph{Is equilibrium always an entropy maximum?}, 
\emph{J.~Stat.~Phys.}~{\bf 35} (1984) 159.
\bibitem{cg1} A.M.~Salzberg, \emph{Exact Statistical Thermodynamics of Gravitational 
Interactions in One and Two Dimensions}, \emph{J.~Math.~Phys.}~{\bf 6} (1965) 158.
\bibitem{cg2} M.E.~Fisher and D.~Ruelle, \emph{The Stability of Many‐Particle Systems}, 
\emph{J.~Math.~Phys.}~{\bf 7} (1966) 260;
\bibitem{cg3} L.G.~Taff, \emph{Celestia Mechanics}, Wiley, New York, (1985) p.~437.
\bibitem{cg4} W.C.~Saslaw, \emph{Gravitational Physics of Stellar and Galactic Systems}, 
Cambridge University Press, Cambridge, (1985) p.~217. 
\bibitem{cg5} D.~Pavon,\emph{Thermodynamics of superstrings}, \emph{Gen.~Rel.~Grav.}~{\bf 19} (1987) 375. 
\bibitem{cg6} J.~Binney and S.~Tremaine, \emph{Galactic Dynamics}, Princeton University Press, 
Princeton, (1987) p.~267. 
\bibitem{cg7} H.E.~Kandrup, \emph{Mixing and "violent relaxation" for the one-dimensional 
gravitational Coulomb gas}, \emph{Phys.~Rev.~A} {\bf 40} (1989) 7265. 
\bibitem{cg8} H.S.~Robertson, \emph{Statistical Thermophysics}, Prentice-Hall, Englewood Cliffs, NJ, (1993) p.~96. 
\bibitem{cg9} H. Bacry, \emph{The existence of dark matter in question}, \emph{Phys.~Lett.~B} {\bf 317} (1993) 523.
\bibitem{Gibbs} J.W. Gibbs, \emph{Elementary Principles in Statistical Mechanics--Developed with Especial Reference 
to the Rational Foundation of Thermodynamics}, C. Scribner's Sons, New York, (1902) (Yale University Press, New Haven, 
1948; OX Bow Press, Woodbridge, Connecticut, 1981), p. 35. 
\bibitem{Tsallis:2012js} C.~Tsallis and L.J.L.~Cirto, \emph{Black hole thermodynamical entropy}, 
\emph{Eur.~Phys.~J.~C} {\bf 73} (2013) 2487.
\bibitem{Davies:1978mf} P.C.W.~Davies, \emph{Proc.~R.~Soc.~Lond.~A},
\emph{Thermodynamics of Black Holes} {\bf 353} (1977) 499.
\bibitem{landsberg1980entropies} P.T.~Landsberg and D.~Tranah, \emph{Entropies need not to be concave},
\emph{Phys.~Lett.~A} {\bf 78} (1980) 219.
\bibitem{bishop1987thermodynamics} N.T.~Bishop and P.T.~Landsberg, \emph{The thermodynamics of a system containing
two black holes and black-body radiation}, \emph{Gen.~Rel.~Grav.} {\bf 19} (1987) 1083.
\bibitem{pavon1986some} D.~Pav\'on and J.M.~Rub\'{\i}, \emph{On some properties of the entropy of a system containing
a black hole}, \emph{Gen.~Rel.~Grav.} {\bf 18} (1986) 1245.
\bibitem{1993Natur.365..103M} J.~ Maddox, \emph{When entropy does not seem extensive}, 
\emph{Nature}, {\bf 365} (1993) 103.
\bibitem{Gour:2003pd} G.~Gour, \emph{Entropy bounds for charged and rotating systems}, 
\emph{Class.~Quant.~Grav.}~{\bf 20} (2003) 3403.
\bibitem{Oppenheim:2002kx} J.~Oppenheim, \emph{Thermodynamics with long-range interactions: 
From Ising models to black holes}, \emph{\pre} {\bf 68} (2003) 016108.
\bibitem{Pesci:2006sb} A.~Pesci, \emph{Entropy of gravitating systems: Scaling laws versus radial profiles}, 
\emph{Class.~Quant.~Grav.}~{\bf 24} (2007) 2283.
\bibitem{Aranha:2008ni} R.F.~Aranha et al., \emph{The Efficiency of Gravitational Bremsstrahlung
Production in the Collision of Two Schwarzschild Black Holes}, \emph{Int.~J.~Mod.~Phys.~D} {\bf 17} (2008) 2049.
\bibitem{Landsberg} P.T.~Landsberg, \emph{Thermodynamics and Statistical Mechanics}, Dover, New York (1990).
\bibitem{Shannon} C.E.~Shannon, \emph{A mathematical theory of communication}, 
\emph{Bell. Syst.~Tech.~J.}~{\bf 27} (1948) 379; ibid.~623.
\bibitem{Khinchin} A.I.~Khinchin, \emph{Mathematical Foundations of Information Theory}, Dover, New York (1957).
\bibitem{Tempesta} P.~Tempesta, \emph{Beyond the Shannon-Khinchin Formulation: The Composability Axiom
and the Universal Group Entropy}, \emph{Ann.~Phys.}~{\bf 365} (2016) 180. 
\bibitem{Tsallis:1987eu} C.~Tsallis, \emph{Possible Generalization of Boltzmann-Gibbs Statistics}, 
\emph{J.~Stat.~Phys.}~{\bf 52} (1988) 479.
\bibitem{Biro:2013cra} T.S.~Bir\'o and V.G.~Czinner, \emph{A $q$-parameter bound for particle spectra based 
on black hole thermodynamics with R\'enyi entropy}, \emph{Phys.~Lett.~B} {\bf 726} (2013) 861.
\bibitem{renyi1959dimension} A.~R\'enyi, \emph{On the dimension and entropy of probability distributions}, 
\emph{Acta.~Math.~Acad.~Sci.~Hung.}~{\bf 10} (1959) 193.
\bibitem{renyi1970probability} A.~Renyi, \emph{Probability Theory}, North Holland, Amsterdam (1970).
\bibitem{Bekenstein:1980jp} J.D.~Bekenstein, \emph{A Universal Upper Bound on the Entropy to Energy Ratio 
for Bounded Systems}, \emph{\prd} {\bf 23} (1981) 287.
\bibitem{Bek1} C.~Beck, \emph{Generalized statistical mechanics of cosmic rays}, \emph{Physica A} {\bf 331} (2004) 173.
\bibitem{Bek2} C.~Beck, \emph{Superstatistics in high-energy physics. Application to 
cosmic ray energy spectra and $e^+e^-$ annihilation}, \emph{Eur.~Phys.~J.~A}{\bf 40} (2009) 267.
\bibitem{BU} T.S.~Bir\'o and K.~\"Urm\"ossy, \emph{Non-extensive equilibration in relativistic matter}, 
\emph{J.~Phys.~G} {\bf 36} (2009) 064044.
\bibitem{Kaburaki:1993ah} O.~Kaburaki, I.~Okamoto and J.~Katz, \emph{Thermodynamic stability of Kerr black holes}, 
\emph{\prd} {\bf 47} (1993) 2234.
\bibitem{Katz:1993up} J.~Katz, I.~Okamoto, and O.~Kaburaki, \emph{Thermodynamic stability of pure black holes},
\emph{Class.~Quant.~Grav.} {\bf 10} (1993) 1323.
\bibitem{kaburaki1996critical} O.~Kaburaki, \emph{Critical behavior of extremal Kerr-Newman black holes}, 
\emph{Gen.~Rel.~Grav.} {\bf 28} (1996) 843.
\bibitem{poincare1885equilibre} H.~Poincar\'e, \emph{Sur l'{\'e}quilibre d'une masse fluide anim{\'e}e d'un 
mouvement de rotation}, \emph{Acta.~Math.} {\bf 7} (1885) 259.
\bibitem{Arcioni:2004ww} G.~Arcioni and E.~Lozano-Tellechea, \emph{Stability and critical phenomena of black 
holes and black rings}, \emph{\prd} {\bf 72} (2005) 104021.
\bibitem{AzregAinou:2012hy} M.~Azreg-A\"{\i}nou and M.~E.~Rodrigues, \emph{Thermodynamical, geometrical 
and Poincar\'e methods for charged black holes in presence of quintessence}, 
\emph{JHEP} {\bf 09} (2013) 146.
\bibitem{Czinner:2015eyk} V.G.~Czinner and H.~Iguchi, \emph{R\'enyi entropy and the thermodynamic stability
of black holes}, \emph{Phys.~Lett.~B} {\bf 752} (2016) 306.
\bibitem{CzM} V.G.~Czinner and F.C.~Mena, \emph{Relative information entropy in cosmology:
The problem of information entanglement}, \emph{Phys.~Lett.~B} {\bf 758} (2016) 9.
\bibitem{Chamblin:1999tk} A.~Chamblin et al., \emph{Charged AdS black holes and catastrophic holography},
\emph{\prd} {\bf 60} (1999) 064018.
\bibitem{Chamblin:1999hg} A.~Chamblin et al., \emph{Holography, thermodynamics and fluctuations of charged
AdS black holes}, \emph{\prd} {\bf 60} (1999) 104026.
\bibitem{Caldarelli:1999xj} M.M.~Caldarelli, G.~Cognola and D.~Klemm, \emph{Thermodynamics of Kerr-Newman-AdS 
black holes and conformal field theories}, \emph{Class.~Quant.~Grav.} {\bf 17} (2000) 399.
\bibitem{Tsai:2011gv} Y-D.~Tsai, X.N.~Wu and Y.~Yang, \emph{Phase Structure of Kerr-AdS Black Hole},
\emph{\prd} {\bf 85} (2012) 044005.
\bibitem{Altamirano:2014tva} N.~Altamirano et al., \emph{Thermodynamics of rotating black holes and black rings:
phase transitions and thermodynamic volume}, \emph{Galaxies} {\bf 2(1)} (2014) 89.
\bibitem{abe2001general} S.~Abe, \emph{General pseudoadditivity of composable entropy prescribed by the 
existence of equilibrium}, \emph{\pre} {\bf 63} (2001) 061105.
\bibitem{Tsallis-bib} http://tsallis.cat.cbpf.br/TEMUCO.pdf.
\bibitem{DNA} C.A.M.~Vald\'es et al., \emph{Nonextensivity and Tsallis entropy in DNA fragmentation 
patterns by ionizing radiation}, \emph{Journal of Modern Physics} {\bf 3} (2012) 431.
\bibitem{okamoto1990thermodynamical} I.~Okamoto and O.~Kaburaki, \emph{The third law of thermodynamics for 
Kerr black holes}, \emph{Mon.~Not.~Roy.~Astron.~Soc.} {\bf 250} (1991) 300.
\bibitem{1978MNRAS.183..765K} J.~Katz, \emph{On the number of unstable modes of an equilibrium}, 
\emph{Mon.~Not.~Roy.~Astron.~Soc.} {\bf 183} (1978) 765.
\bibitem{1979MNRAS.189..817K} J.~Katz, \emph{On the Number of Unstable Modes of an Equilibrium - Part Two}, 
\emph{Mon.~Not.~Roy.~Astron.~Soc.} {\bf 189} (1979) 817.
\bibitem{Sorkin:1981jc} R.~Sorkin, \emph{A Criterion for the onset of instability at a turning point},
\emph{Astrophys.~J.}~{\bf 249} (1981) 254.
\bibitem{1994PhLA..185...21K} O.~Kaburaki, \emph{Should entropy be concave?}, 
\emph{Physics Letters A} {\bf 185} (1994) 21.
\bibitem{czinner2015black} V.G.~Czinner, \emph{Black hole entropy and the zeroth law of thermodynamics}, 
\emph{Int.~J.~Mod.~Phys.~D} {\bf 24} (2015) 1542015.
\bibitem{CzIg} V.G.~Czinner and Hideo Iguchi, \emph{A zeroth law compatible model to Kerr black hole thermodynamics}, 
\emph{Universe} {\bf 3} (2017) 14.
\bibitem{Biro:2012bka} T.S.~Bir\'o, \emph{Ideal gas provides q-entropy}, \emph{Physica A} {\bf 392} (2013) 3132.
\bibitem{Biro:2013qea} T.S.~Bir\'o et al., \emph{Quark-gluon plasma connected to finite heat bath}, 
\emph{Eur.~Phys.~J.~A} {\bf 49} (2013) 110.
\bibitem{Carlip:2014pma} S.~Carlip, \emph{Black Hole Thermodynamics}, 
\emph{Int.~J.~Mod.~Phys.~D} {\bf 23} (2014) 1430023.
\end{thebibliography}
\end{document}